\documentclass[11pt]{article}

\usepackage{geometry}
\usepackage[numbers]{natbib}
\usepackage{amsfonts,amsmath,amssymb,amsthm}
\usepackage{graphicx}
\usepackage[colorlinks,citecolor=green,linkcolor=blue,bookmarks=false,hypertexnames=true]{hyperref}
\usepackage[sf,bf]{titlesec}
\usepackage[color=yellow]{todonotes}
\usepackage{lettrine}

\newcommand{\defn}[1]{\emph{\textbf{#1}}}

\allowdisplaybreaks
\usepackage{titling}
\usepackage{array}
\preauthor{\begin{center}
    \large \lineskip .75em%
        \begin{tabular}[t]{>{\centering\arraybackslash}p{.45\textwidth}}}
        \postauthor{\end{tabular}\par\end{center}}
\makeatletter
\renewcommand\and{% \begin{tabular}
  \end{tabular}%
  \hfill
  \begin{tabular}[t]{>{\centering\arraybackslash}p{.45\textwidth}}}% \end{tabular}
\makeatother

\usepackage{framed}
\definecolor{shadecolor}{RGB}{204,229,255}

\begin{document}

%\title{A Markov Chain Assessment of the Weak Impermeability of the Eastern Pacific Barrier}
\title{Breaching the Barrier: Transition Pathways of Coral Larval Connectivity Across the Eastern Pacific}

\author{M.J.\ Olascoaga\thanks{Corresponding author.}\\ Department of Ocean Sciences\\ Rosenstiel School of Marine, Atmospheric \& Earth Science\\ University of Miami\\ Miami, Florida, USA\\ \href{mailto:jolascoaga@miami.edu}{\texttt{jolascoaga@miami.edu}} \and F.J.\ Beron-Vera\\ Department of Atmospheric Sciences\\ Rosenstiel School of Marine, Atmospheric \& Earth Science\\ University of Miami\\ Miami, Florida, USA\\ \href{mailto:fberon@miami.edu}{\texttt{fberon@miami.edu}} \and G.\ Bonner\\ Morgridge Institute for Research\\ 330 N Orchard St\\ Madison, WI 53715, USA\\ \href{mailto:gbonner@morgridge.org}{\texttt{gbonner@morgridge.org}} \and C.\ McKean\\ Department of Biological Science\\ University of Pittsburgh\\ Pittsburgh, Pennsylvanian, USA \href{mailto:com180@pitt.edu}{\texttt{com180@pitt.edu}}} 

\date{Started: October 31, 2025.  This version: \today.}

\maketitle

\newpage

\begin{abstract}
  Genetic analyses indicate minimal gene flow across the so-called Eastern Pacific Barrier (EPB) in larvae of the reef-building coral \emph{Porites lobata}. Notably, Clipperton Atoll, situated on the eastern side of the EPB, is the only site that exhibits detectable genetic connectivity with the Line Islands, which lie to the west of the EPB. To elucidate the relationship between this genetic signal and large-scale Pacific Ocean circulation, we analyze historical trajectories of surface-drifting buoys from the Global Drifter Program (GDP). We first discretize the GDP drifter trajectories into a Markov chain representation and subsequently apply transition path theory (TPT) in combination with Bayesian inference. The TPT analysis identifies reactive trajectories---pathways that connect the Line Islands to Clipperton Atoll with minimal detours---whose travel times do not exceed 5 months, which is taken as an upper bound for the larval survival time of \emph{P. lobata}. Consistently, the posterior distribution of transport from Pacific islands west of the EPB to Clipperton Atoll attains a local maximum in the Line Islands at a travel time of approximately 2.5 months. Our probabilistic characterization of the Lagrangian dynamics therefore supports a scenario of weak, but non-negligible, permeability of the EPB, in agreement with the genetic evidence, and it motivates a refined dynamical definition of the EPB based on the remaining duration of reactive trajectories. Furthermore, our results indicate that the connectivity between the Line Islands and Clipperton Atoll is governed primarily by the seasonal modulation of the North Equatorial Countercurrent, rather than by the phase of the El Ni\~no--Southern Oscillation (ENSO). Finally, Clipperton Atoll's role as a terminal sink for trajectories is relevant to the planned mining operations.
\end{abstract}

\noindent Eastern Pacific Barrier $\mid$ Coral-reef building larvae $\mid$ Markov chain $\mid$ TPT $\mid$ Bayesian inversion

\newpage

\begin{snugshade*}

\section{Significance Statement}

The Eastern Pacific Barrier (EPB) is usually seen as a major obstacle to the spread of reef‑building corals, but genetic data show an unexpected link between Clipperton Atoll on the eastern side and the Line Islands on the western side. Using 30 years of data from drifting ocean buoys together with transition path theory for Markov chains and Bayesian inference, we find that rare but realistic ocean currents can carry coral larvae from the Line Islands to Clipperton within their likely 5‑month lifetime. These special transport routes are mainly driven by the seasonal strengthening of the North Equatorial Countercurrent, rather than by El Niño events. Our findings offer a clear physical explanation for the observed genetic connection across the EPB, provide a new way to define this long‑standing biogeographic barrier, and show how tracking ocean flows can improve predictions of how marine populations are connected as the climate changes. The role of Clipperton Atoll as sink of trajectories bears significance for planned mining.\\% >120 words, plain

\begin{quote}
    \flushright{\emph{\textup{``}The best preparation for tomorrow is doing your best today.\textup{''}}}

    \vspace{-1em}
        
    \flushright{H. Jackson Brown Jr., Life's Little Instruction Book}
\end{quote}

\end{snugshade*}

\newpage

%\tableofcontents\bigskip

\newpage

\noindent \lettrine[findent=2pt]{\fbox{\textbf{T}}}{ }he extensive stretch of deep ocean that separates the central from the eastern Pacific has long been hypothesized to constitute a major barrier to larval dispersal, with such ideas traceable at least to \citet{Darwin-59} and \citet{Ekman-53}. In Chapter XI, p.\@~348, Darwin characterized this region as ``a barrier of another type,'' describing it as ``impassable'' and as keeping ``totally distinct fauna(s)'' apart, whereas Ekman described it as ``the world’s most potent marine barrier'' to larval dispersal. Ekman designated this region the \emph{Eastern Pacific Barrier} (\emph{EPB}).

In line with Darwin’s and Ekman’s idea of an effective EPB, the Central Pacific Ocean---west of the EPB---supports a high richness of coral species, with more than 100 distinct species thriving on its reefs. In contrast, the Eastern Tropical Pacific---east of the EPB---exhibits very low coral diversity, containing only three dominant reef-building species throughout the region \cite{Glynn-97}. The species-poor state of coral communities in the Eastern Tropical Pacific is attributed to their geographic isolation from regions with high species richness and to the frequent environmental disturbances that affect this area \cite{Glynn-96}. Genetic analyses indicate that the reef-building coral \emph{Porites lobata} exhibits very weak gene flow across the EPB \cite{Baums-etal-12}. The small leakage is attributed to Clipperton Atoll, which stands out as a notable exception: although it lies on the eastern side of the EPB, it shows genetic connectivity with the Line Islands to the west. Using Bayesian clustering analyses, \cite{Baums-etal-12} treated five Clipperton samples (from the Eastern Tropical Pacific) as unknowns and attempted to assign them to three Pacific regions (Central/Eastern and Hawaii). Under an admixture model, these samples had a higher mean assignment probability to the Central Pacific than to the Eastern Tropical Pacific, with only negligible assignment to Hawaii.

However, the apparent effectiveness of the EPB seems inconsistent with the prevailing circulation in the Pacific. The equatorial current system is in fact dominated by westward flows \cite{Kessler-06, Wyrtki-etal-81}, so connectivity would be expected primarily from east to west. The eastward-flowing surface North Equatorial Countercurrent (NECC) \cite{Wyrtki-Kendall-67} is the only mechanism that can facilitate dispersal toward the eastern Pacific \cite{Richmond-90}. The NECC weakens downstream as water subducts into the thermocline and flows toward the equator, with linear Sverdrup theory \cite{Sverdrup-47} providing a first-order explanation for it as a geostrophic flow driven by meridional variations in wind stress curl associated with the off-equatorial Intertropical Convergence Zone (ITCZ). Nonlinear advection accounts for the interaction of the NECC with mesoscale eddies \cite{McPhaden-Ripa-90, McPhaden-96a} and plays a central role in determining the large-scale structure of the current system \cite{Kessler-etal-03}, which exhibits variability on both seasonal \cite{Philander-etal-87, Hsin-Qiu-12a} and interannual time scales associated with the El Ni\~no–Southern Oscillation (ENSO) \cite{Wyrtki-73b, Hsin-Qiu-12b}.

The aim of this study is to examine the connectivity of larvae of the reef‑building coral \emph{P.\ lobata}, commonly known as lobe coral, across the EPB from a physical oceanographic standpoint by evaluating their Lagrangian transport governed by the equatorial current system in the Pacific Ocean. This analysis is carried out under the assumption that larvae are passively transported by surface‑ocean currents while also experiencing diffusive dispersal, which provides a basic representation of their free‑swimming capability as a stochastic process. The methodology adopted in this study employs trajectories of satellite-tracked surface drifting buoys as proxies for larval dispersal pathways. Preliminary research \cite{McKean-22} focusing on individual trajectories has offered initial insights into cross-EPB connectivity and its temporal variability. Here, we present an inherently probabilistic framework that concentrates on studying trajectory distributions, which capture statistically robust and dynamically meaningful properties of the underlying Lagrangian flow. To facilitate this focus on such distributions, we first represent trajectory motion as a Markov chain via an appropriate discretization. A Markov chain is a mathematical model for a sequence of random events in which the next outcome depends only on the current state, and changes over time follow a probabilistic rule that assigns chances of moving between states \cite{Norris-98, Bremaud-99, Privault-18}.

We then employ two specialized techniques. First, we use transition path theory (TPT) \cite{E-VandenEijnden-06, VandenEijnden-06, Metzner-etal-09}, which enables us to rigorously and statistically characterize ensembles of trajectories that connect source and target with minimal detours, the so‑called reactive trajectories. In our TPT application, the target domain naturally comprises Pacific Ocean islands and coastal continental regions located east of the EPB, while the source domain consists of a set of islands situated west of the EPB. This configuration is intended to represent plausible source regions of \emph{P.\ lobata} larvae observed east of the EPB, particularly at Clipperton Atoll, which the analysis finds to emerge as an attractor of trajectories originating west of the EPB, specifically the Line Islands. Second, we employ Bayesian inversion \cite{Bolstad-Curran-16} as a complementary approach to infer the geographic origin of larvae observed at Clipperton Atoll. The TPT analysis indicates that the EPB is only partially impermeable, in agreement with the genetic analysis, and it motivates a refined definition of the EPB grounded in a TPT statistic that describes the remaining duration of reactive trajectories.

Markov chain reduction applied to drifter trajectories has already been used since the early 2010s \cite{Maximenko-etal-12, vanSebille-etal-12}. The application of TPT in oceanography is comparatively recent \cite{Miron-etal-21-Chaos, Miron-etal-22, Beron-etal-23-JPO, Beron-etal-22-ADV}, as is the use of Bayesian inversion techniques \cite{Miron-etal-19-Chaos, Beron-etal-26-PNAS}. Beyond the particular methodological tools adopted, we emphasize the observation-based character of our approach, which stands in contrast to numerical modeling studies of larval connectivity across the EPB \cite{Wood-etal-14, Wood-etal-16}.

%The remainder of the paper follows a standard structure. Section 2 describes the data (Section 2.1) and reviews the basic results of Transition Path Theory (TPT) for Markov chains, as well as the Bayesian inference framework adopted in the Markov chain setting (Section 2.2). Additional methodological details are provided in the Supplementary Material. Section 3 presents the main results. Section 4 discusses the implications of the drifter-based findings for the connectivity of \emph{P. lobata}, including an examination of the influence of seasonal variability (Section 4.1) and El Ni\~no–Southern Oscillation (ENSO) phases (Section 4.2). Finally, Section 5 summarizes the main conclusions.

\section{Methodology}

\subsection{Data}

Our data are obtained from the NOAA Global Drifter Program (GDP) \cite{Lumpkin-Pazos-07}, which has provided surface drifting buoy trajectories tracked by satellites since 1979, using the \emph{Argos} system or the Global Positioning System (GPS). The drifters in this database consist of a spherical surface float coupled to a 15-m-long holey-sock drogue; they are specifically designed to minimize wind slippage and wave-induced drift, thus allowing the instruments to follow the ambient near-surface water motion as faithfully as possible \cite{Sybrandy-Niiler-91, Niiler-Paduan-95}. The drogue may be lost along a drifter trajectory \cite{Lumpkin-etal-12}. Unlike earlier work \cite{Miron-etal-19-Chaos, Miron-etal-21-Chaos, Beron-etal-22-ADV, Bonner-etal-23}, which applied techniques similar to those used in this study in the context of marine debris and macroalgal dispersal, we restrict our attention to those portions of trajectories during which the drogue is present, as their motion is expected to more closely represent that of larvae of reef-building coral \emph{Porites lobata}, which are transported by ocean currents and are less affected by windage and wave motion than floating matter such as plastic litter and pelagic \emph{Sargassum} seaweed.

\subsection{Background Probability Theory}

The probabilistic methods used to analyze the GPD data assume that the concentration of larvae has been measured in a region $B$ of the world ocean domain $D$. It is further assumed that there is another region $A$, disjoint from $B$, which is hypothesized to be the origin of the larvae. A key working hypothesis, dictated primarily by the characteristics of the available data, is that the larvae are advected as passive tracers by temporally uniform currents while simultaneously undergoing diffusive spreading, which minimally accounts for larval motility.

\subsubsection{Markov Chain Reduction}

The starting point of the probabilistic techniques is to model the motion of the GDP drifters as a Markov chain, that is, a stochastic model of state transitions in which the transition probability of each state depends only on the state attained in the previous step \cite[cf.][or any standard textbook on Markov chains]{Norris-98, Bremaud-99, Privault-18}. A time-homogeneous, discrete Markov chain arises when evolution under an advection–diffusion equation with steady drift in $D$ is discretized via Ulam's method \cite{Ulam-60}. This involves introducing the notion of a transfer operator defined using a stochastic kernel, which is elaborated in \cite{Lasota-Mackey-94}, while the Ulam discretization method applied in this context is worked out in full detail in \cite{Miron-etal-19-JPO, Miron-etal-21-Chaos}.

Let $f(x) \ge 0$ such that $\int_D f(x)\,dx = 1$ be a probability density on $D$---thought of as a space where we can talk about probabilities, using a special collection of subsets for which areas (and thus probabilities) are defined---at any given time. In our case, this represents the normalized concentration of larvae in $D$, at any given time. Let $\mathcal P$ denote the transfer operator, which evolves $f(x)$ to a new distribution $(\mathcal P f)(y) = \int_D f(x) k(x,y)\,dx$ after a time interval of length $\Delta t$, where $k(x,y) \ge 0$ with $\int_D k(x,y)dy = 1$ is a stochastic kernel.

The discretization method consists in projecting $f$ and $\mathcal P$ onto a finite-dimensional vector space spanned by characteristic functions indicating membership in each cell of a partition $C_1, C_2, \dotsc, C_N$ of $D$. In this way, $f$ is reduced to a probability vector $\mathbf f \in \mathbb R^{1 \times N}$, i.e., satisfying $\smash{\sum_{i=1}^N} f_i = 1$, supported on the cells of the partition, and $\mathcal P$ is reduced to a stochastic matrix $P \in \mathbb R^{N \times N}$ whose $(i,j)$th entry gives the conditional probability of transitioning in one time step ($\Delta t$) between cells $i$ and $j$, which from the states of the Markov chain. That is,
\begin{align}
    P_{ij} := \mathbb P(X_{n+1} = j \mid X_n = i),
    \label{eq:P}
\end{align}
where $X_n$ denotes the random position within $D$ at the discrete time instant $n \Delta t$, with $D = \bigcup_{i=1}^N C_i$ identified with the discrete index set $\{1,2,\dotsc,N\}$. The latter is endowed with the probability measure $\mathbb P$ and is thereby regarded as the underlying probability space.

Using the drifter observations, the transition probability $P_{ij}$ in \eqref{eq:P} can be estimated by counting the number of drifters that occupy cell $C_i$ at any time $t$ and are subsequently found in cell $C_j$ at time $t + \Delta t$, and then normalizing this count by the total number of drifters present in $C_i$ at time $t$. This procedure yields a transition matrix $P$ that is row-stochastic, i.e., $\smash{\sum_{j=1}^N} P_{ij} = 1$ for $i = 1,2,\dotsc,N$. The probability vector $\mathbf f$ then evolves after a time interval of length $\Delta t$ by left multiplication by $P$, that is, to $\mathbf f P$. 

We assume that there exists a probability vector $\boldsymbol\pi$, called a stationary distribution, which is invariant (i.e., $\boldsymbol\pi P = \boldsymbol\pi$) and is reached from any initial distribution $\mathbf f$ in the limit of infinitely many applications of $P$. The existence of $\boldsymbol\pi$ is guaranteed under the condition that $P$ is irreducible and aperiodic. We refer the reader to standard textbooks on Markov chains for formal definitions and only note that these conditions imply that every state of the Markov chain is visited, independently of the starting state, and that no state is visited cyclically. Beyond this ergodicity assumption, we further assume that the chain is in stationarity; that is, the probability that the chain is in state $i$---a larva happens to lie in cell $C_i$ of the partition of the flow domain $D$---is $\pi_i$.

\subsubsection{Transition Path Theory}

The transition path theory (TPT) developed in \cite{E-VandenEijnden-06, VandenEijnden-06, Metzner-etal-09} provides a rigorous probabilistic characterization of the ensemble of paths that run from region $A \subset D$ to region $B \subset D$ without returning to $A$ or passing through $B$. Connecting $A$ and $B$ with minimal detours, such \emph{reactive trajectories} facilitate transport from $A$ to $B$ in a highly efficient manner and are therefore of particular interest, especially when this transport involves biological tracers that are subject to natural mortality, such as larvae. Consequently, TPT is expected to provide insight into the origin of larvae in $B$ as $A$ can be defined as the union of possibly disconnected of the partition of the domain $D$ of the flow.

The probabilistic characterization of reactive trajectories is carried out using various statistics computable from the transition matrix $P$, the stationary distribution $\boldsymbol\pi$, and the forward ($\mathbf q^+$) and backward ($\mathbf q^-$) committors, with $\mathbf q^+$ giving the probability of first entering $A$ rather than $B$, and $\mathbf q^-$ that of last exiting $B$ rather than $A$. They can be computed, as reviewed in the Supplementary Material, by solving systems of linear algebraic equations involving $P$ and $P^-$, respectively. Here, $P^-_{ij} = \pi_j P_{ji} / \pi_i$ is the transition matrix of the chain traversed in reverse time. Explicit formulas for these forward and backward committor probabilities \cite{Metzner-etal-09, Helfmann-etal-20} are given in the Supplementary Material. Of particular significance are the following TPT statistics (explicit formulas are reviewed in the Supplementary Material):
\begin{itemize}
   \item the \emph{density of reactive trajectories}, $\boldsymbol\mu^{AB}$, defined as the normalized joint probability that the chain is in state $i$ while transitioning from $A$ to $B$, which indicates where trajectories spend most of their time \cite{Metzner-etal-09, Helfmann-etal-20}; 
   \item the \emph{current of reactive trajectories}, $f^{AB}$, which represents the net average flux of trajectories passing through states $i$ and $j$ at two consecutive times while on a direct route from $A$ to $B$ \cite{Metzner-etal-09, Helfmann-etal-20}; 
   \item the \emph{arrival rate}, $k^{i \in B \leftarrow}$, computed by summing the reactive flux entering a particular state $i \in B$, which is the probability per time step that a reactive trajectory leaves $i \in A$, rather than the set $B$ as a whole, as discussed in \cite{Metzner-etal-09, Helfmann-etal-20}, and an extension, which we refer to as the \emph{departure rate}, $k^{i \in A \to}$, computed by summing the reactive flux exiting a particular state $i \in A$, which is the probability per time step that a reactive trajectory leaves $i \in A$, rather than the set $A$ as a whole; and
   \item the \emph{remaining duration of reactive trajectories}, $t^{iB}$, that is, the expected first entrance time to $B$ from any state $i$ after leaving $A$ on a direct route to $B$ \cite{Bonner-etal-23}.
\end{itemize}

\subsubsection{Bayesian Inversion}

 Ignoring how trajectories connect $A$ with $B$, one can seek to infer the posterior probability distribution $p(a \mid t^B)$ for the origin $a \in A$ of the larva concentration observed in $B$ after $t^B$, i.e., some given multiple of time steps of length $\Delta t$. 

This inference is carried out via a Bayesian inversion \cite{Bolstad-Curran-16} on the underlying Markov chain, following the approach of \cite{Miron-etal-19-Chaos} used to reconstruct the likely crash site of an airplane. Bayes' theorem yields $p(a\!\mid\!t^B) \propto p(t^B\!\mid\!a)p(a)$ where $p(a)$ is the prior distribution of $a$, representing knowledge about the origin before any observations are made. Here, $p(t^B\!\mid\!a)$ denotes the posterior probability that the larva is in $B$ after $t^B$ units of time, given that the Markov chain is initialized at $a \in A$.

As detailed in the Supplementary Material, the posterior probability $p(t^B\!\mid\!a)$ can be computed by initializing a probability vector concentrated at state $a$, propagating this vector forward under left multiplication by the transition matrix $P$, and recording the probability mass that accumulates in the set $B$ over time.

Finally, we consider a variation of the above setting in which the sets $A$ and $B$ are directly connected, as prescribed by TPT. Note that the Bayesian inference procedure involves only forward-time evolution; hence, it is the forward reactivity that is relevant. To incorporate this feature into the Bayesian framework, it is appropriate to replace the transition matrix $P$, defined in \eqref{eq:P}, with the matrix $P^+$ with $(i,j)$th entry defined by
\begin{align}
    P^+_{ij} := \mathbb P(X_{n+1}=j\mid X_n = i,\tau^+_B(n + 1)<\tau^+_A(n + 1)) = \frac{P_{ij}q^+_j}{\sum_{j\in D} P_{ij}q^+_j}.
    \label{eq:P+}
\end{align}
Here, $X_n$ denotes the random position in $D$ at time $n\Delta t$, and $\tau^+_A(n+1)$ and $\tau^+_B(n+1)$ denote the random time since $(n+1)\Delta t$ of first entrance into the sets $A$ and $B$, respectively. A formal definition of these random variables is provided in the Supplementary Material. The matrix $P^+$ is row-stochastic, and its $(i,j)$th entry corresponds to the conditional probability of being in state $j$, given that the system was in state $i$ one step earlier (i.e., $P_{ij}$), under the additional condition that a trajectory initiated in $A$ is committed to reach $B$ (as opposed to returning to $A$) in forward time. We call $P^+$ a \emph{forward-reactive transition matrix}. For a detailed derivation of its formula \eqref{eq:P+}, which is employed in a different context than Bayesian inversion with committor constraint, the reader is referred to Appendix C of \cite{Beron-etal-26-JAS}.

\subsection{Setup}

\begin{figure}[t!]
    \centering
    \includegraphics[width=\linewidth]{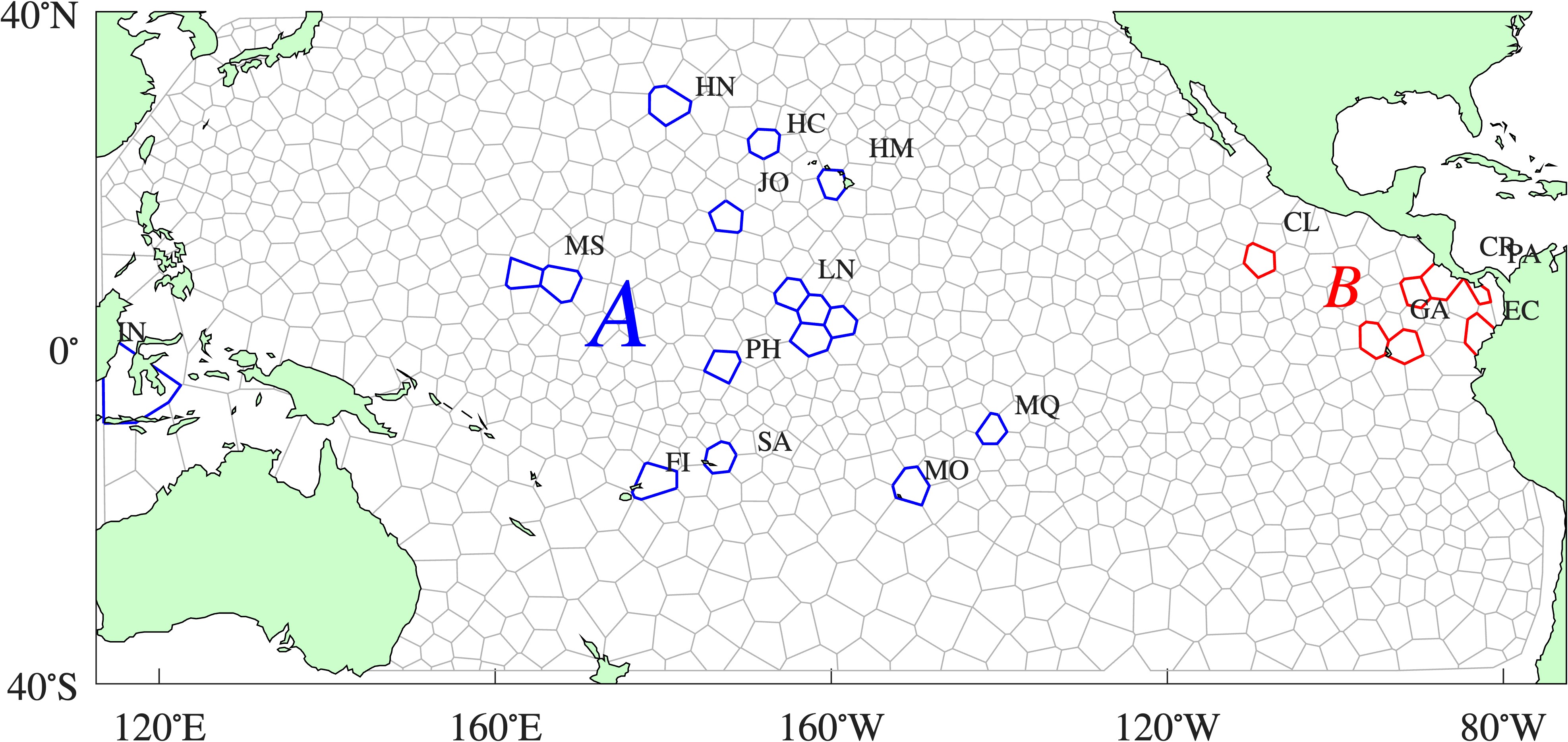}
    \caption{Partition of the Pacific Ocean domain of interest into Voronoi cells resulting from $k$-means clustering of trajectories of satellite-tracked, drogued surface drifting buoys from the NOAA Global Drifter Program (GDP). Indicated are the cells that form the source $A$ and target $B$ for the transition path theory (TPT) calculation. See text for acronyms and details.}
    \label{fig:partition}
\end{figure}

For the purposes of this study, we consider a rectangular domain of the Pacific Ocean extending approximately from $110^\circ$E to $70^\circ$W in longitude and from $40^\circ$S to $40^\circ$N in latitude (Figure \ref{fig:partition}). We denote this region by $D^*$. Within $D^*$, the NOAA GDP dataset contains observations from 26,624 drogued drifters over the period from 15 February 1979 to 20 January 2025. 

The domain $D^*$ is partitioned into nonrectangular cells obtained via a Voronoi tessellation based on the geographical positions of all drifters, with the Voronoi generators determined by applying $k$-means clustering to these drifter locations. Note that the resulting Voronoi cells do not fully cover $D^*$, but instead span a somewhat smaller region, which defines the domain $D$ used for the formulation of the Markov chain model. As discussed in the Supplementary Material, and consistent with previous Markov chain reductions of GDP trajectory data and related datasets \cite{Miron-etal-17, Miron-etal-19-JPO, Miron-etal-22, Beron-etal-23-JPO}, this partial spatial coverage is employed to strengthen communication among the states of the chain and thereby promote approximate ergodicity, a property required for the rigorous application of TPT.

To construct the transition matrix $P$, we sample the trajectories at uniform time intervals of length $\Delta t = 5$ days. The counting of transitions of duration $\Delta t = 5$ days among the cells in $D$ yields a substochastic matrix $P$. In order to apply TPT, $P$ is rendered stochastic by suitably augmenting the Markov chain via the introduction of a virtual state that absorbs probability imbalances and reinjects them into the chain using trajectory information outside $D$, as proposed in \cite{Miron-etal-21-Chaos} and discussed in detail in the Supplementary Material. We emphasize the distinction between $D^*$ and the domain $D$ covered by the Voronoi cells that define the Markov chain.  

The selected temporal increment of $\Delta t = 5$ days exceeds the characteristic Lagrangian decorrelation timescale, which is on the order of 1 day at the ocean surface \cite{LaCasce-08}. This choice ensures a sufficiently pronounced loss of memory of prior states for the Markovian assumption to be approximately satisfied. Comparable temporal resolutions have been adopted in Markov chain–based model reductions of drogued GDP trajectory data \cite{Maximenko-etal-12, vanSebille-etal-12, Beron-etal-20-Chaos, Beron-etal-22-ADV, Drouin-etal-22}. 

Moreover, employing Voronoi cells produces temporal TPT statistics that are largely robust to the particular choice of $\Delta t$ and to the total number of cells, as shown in \cite{Bonner-etal-23}. In this work, we use $10^3$ cells in total, each with an average equivalent rectangular size of 3.5$^\circ$ by 3.5$^\circ$.

The definition of the source set $A \subset D$ and the target set $B \subset D$ for the TPT analysis is informed by empirical observations of the reef-building coral \emph{P.\ lobata}, as reported in \cite{Baums-etal-12}. Population genetic analyses indicate weak gene flow across the EPB in \emph{P. lobata}, with Clipperton Atoll---situated on the eastern side of the EPB---constituting the only location exhibiting genetic connectivity with the Line Islands on the western side of the EPB.

We restrict our analysis to those island and continental coastal locations shown in the map in the top panel of Figure~1 in \cite{Baums-etal-12} where \emph{P.\ lobata} was sampled and that spatially intersect the cells of our partition of the Pacific Ocean domain. Several of these locations, hereafter referred to collectively and loosely as ``islands,'' are too small to be explicitly resolved on our grid; consequently, only a subset of the islands examined in \cite{Baums-etal-12} is included in the present study. The cells that intersect the sampled locations are depicted in Figure~\ref{fig:partition}. Some cells are annotated with a single acronym. From west to east, these acronyms are: IN (Indonesia); MS (Marshalls); FI (Fiji); SA (Samoa); HN (Hawaii North); PH (Phoenix Islands); JO (Johnston Atoll); HC (Hawaii Central); LN (Line Islands); HM (Hawaii Main); MO (Moorea); MQ (Marquesas); CL (Clipperton Atoll); GA (Galapagos); EC (Ecuador); CR (Costa Rica); and PA (Panama).

We define the set $A$ as the disjoint union of all labeled cells situated to the west of the EPB, which we approximate by the meridional band bounded by 140$^\circ$W and 120$^\circ$W, and the set $B$ as the disjoint union of all labeled cells located to the east of the EPB. This definition is modified when applying Bayesian inference. The modification is informed by the results of the TPT analysis, and thus the redefinition is deferred until this analysis is applied. We anticipate, however, that such results will be consistent with genetic-based connectivity inferences for \emph{P. lobata} \cite{Baums-etal-12}.

\section{Results}

\begin{figure}[t!]
    \centering
    \includegraphics[width=.49\linewidth]{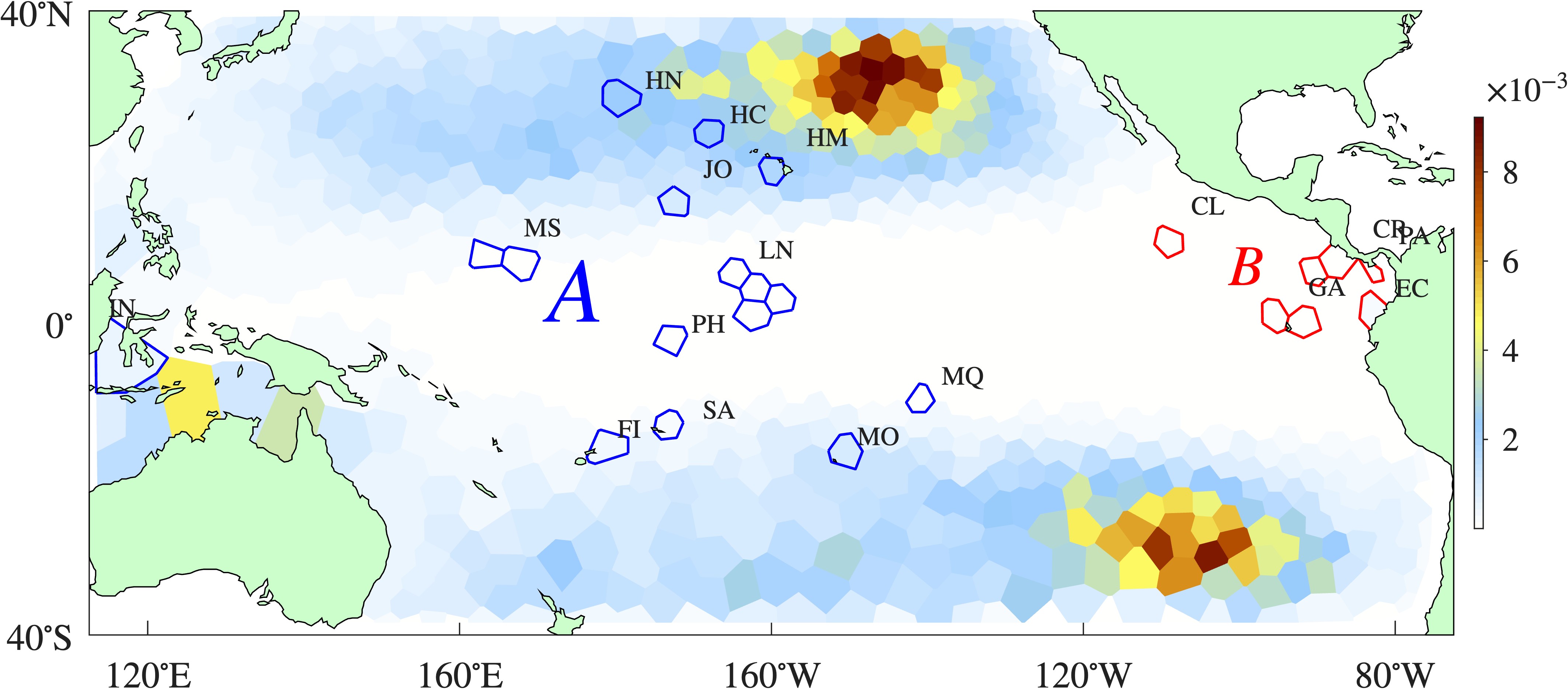}\,
    \includegraphics[width=.49\linewidth]{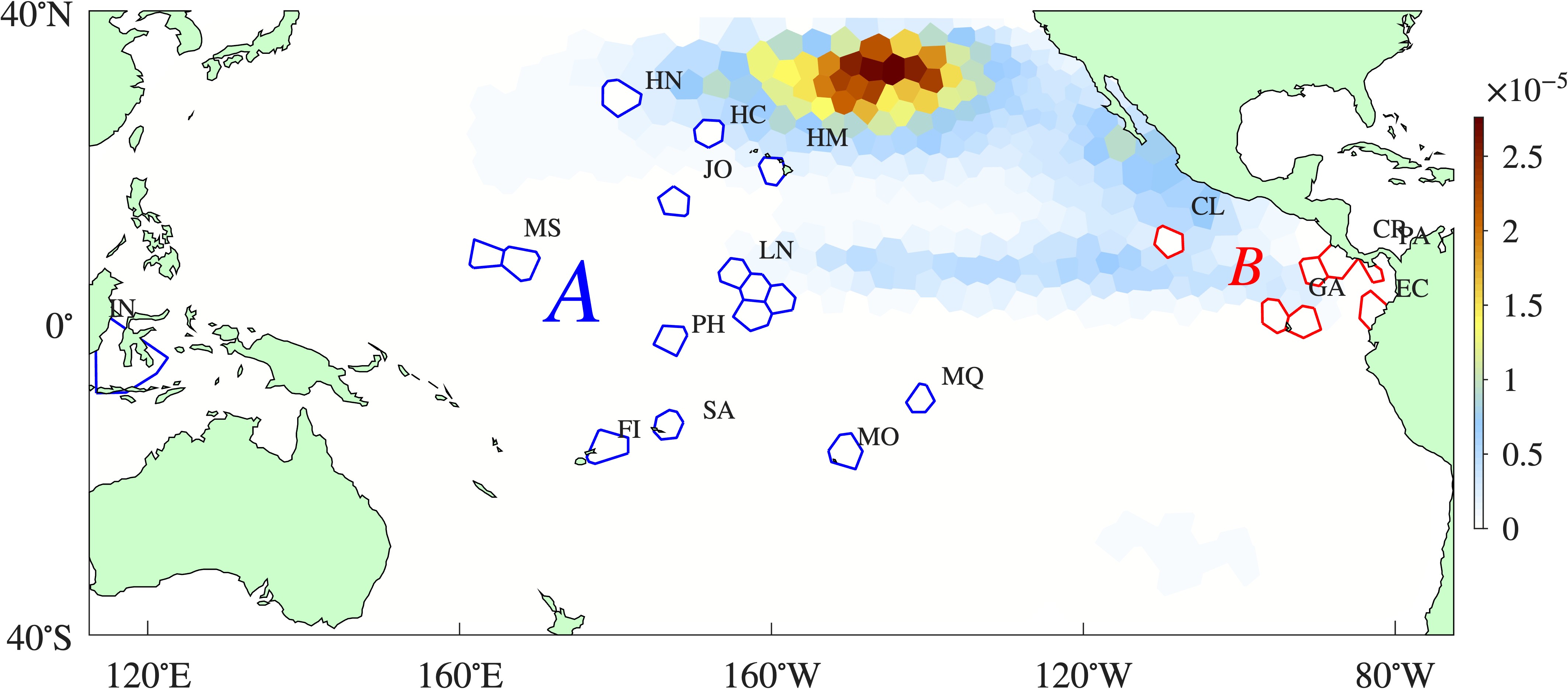}
    \caption{(left panel) Stationary distribution of the Markov chain on cells, constructed via Ulam's discretization method and applied to GDP trajectories assumed to be produced by a stationary stochastic process, showing where the trajectories accumulate in the long run. (right panel) Density of reactive trajectories or normalized joint probability that the chain is in state (cell) $i$ while transitioning from $A$ to $B$, indicating where trajectories spend most of their time.}
    \label{fig:stationary}
\end{figure}

We begin by examining the left panel of Figure~\ref{fig:stationary}, which depicts the stationary distribution $\boldsymbol\pi$ associated with the Markov chain constructed from the GPD drifter trajectories. To be precise, this figure shows $\boldsymbol\pi$ restricted to $D$, the physical Pacific Ocean domain of definition of the Markov chain. This distribution exhibits two local maxima in $D$ positioned near the centers of the subtropical gyres in the Pacific. An arbitrarily chosen initial distribution of larvae over $D$ evolves under the dynamics of the constructed Markov chain and, in the long-time limit (i.e., asymptotically as time tends to infinity), converges to the stationary distribution $\boldsymbol{\pi}$, should, of course, the larvae live indefinitely. As expected, under stationary conditions the local maxima of $\boldsymbol\pi$ align with the well-known regions of plastic debris accumulation \cite{Cozar-etal-14}, particularly in the Northern Hemisphere, the so-called garbage patches associated with wind-stress–curl-driven Ekman transport dynamics \cite{vanSebille-etal-12, Maximenko-etal-12}. Effects of inertia, which are known to accelerate the development of the garbage patches \cite{Beron-etal-16, Beron-etal-19-PoF, Espinosa-Colmenares-26}, likely play a smaller role here because the drifters are drogued.

\begin{figure}[t!]
    \centering
    \includegraphics[width=\linewidth]{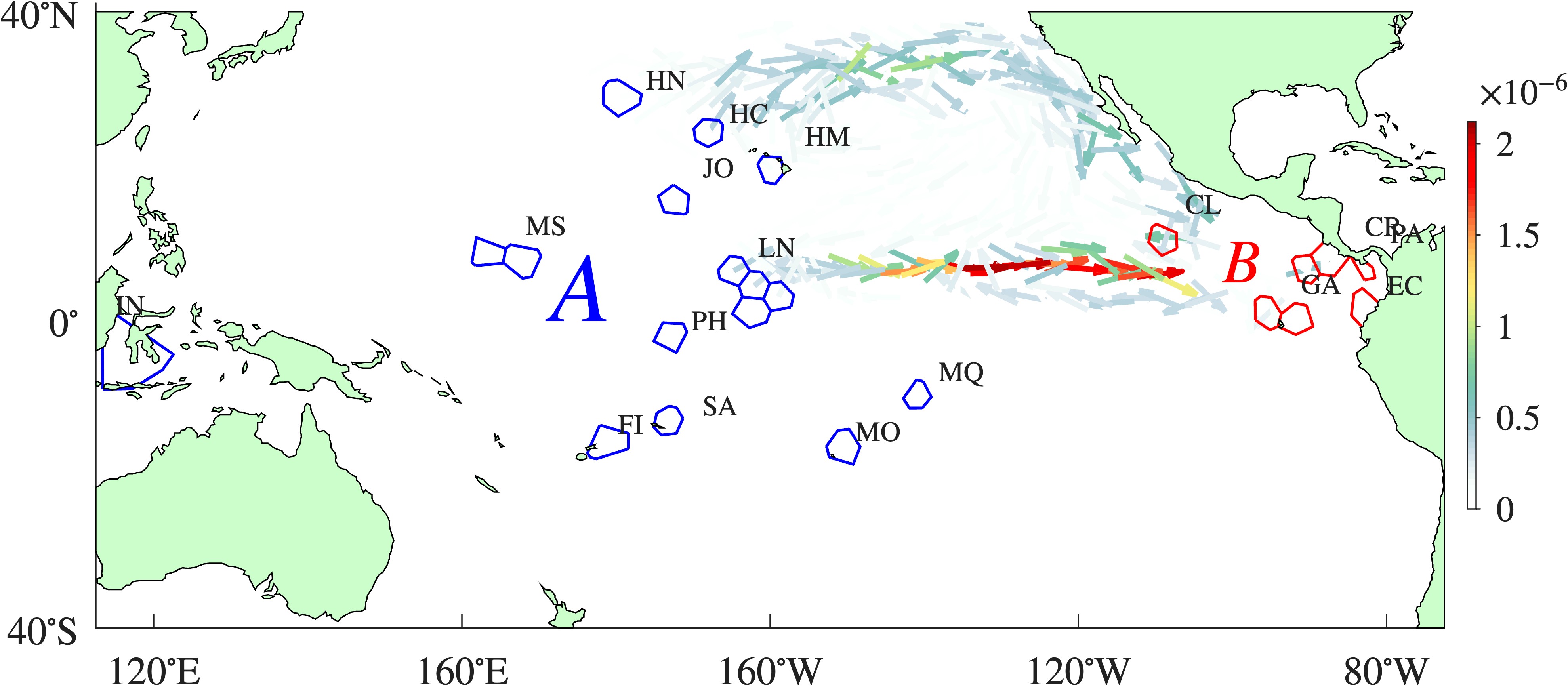}
    \caption{Reactive currents representing the net average flux of trajectories passing through states $i$ and $j$ at two consecutive times while on a direct route from $A$ to $B$.}
    \label{fig:currents}
\end{figure}

Together with the probabilities of forward commitment to the target set $B\subset D$ and backward commitment to the source set $A\subset D$ in the TPT framework, the stationary distribution determines the ensemble of reactive currents that connect $A$ and $B$ with minimal detours, as shown in Figure~\ref{fig:currents}. More precisely, the figure shows the \emph{effective} reactive current $f^+ := (\max\{f^{AB}_{ij} - f^{AB}_{ji},0\})_{i,j\in D}$, which gives the net amount of reactive current passing consecutively through cells $i$ and $j$. This is visualized by attaching to each $i\in D$ the vector $\sum_{j\neq i} f^+_{ij}\mathbf e_{ij}$, where $\mathbf e_{ij}$ is the unit vector pointing from the center of cell $i$ to the center of cell $j$. Note the pronounced reactive current that originates in the cells designated LN, which are geographically associated with the Line Islands and are elements of set $A$, and then flows toward set $B$. As it progresses eastward, it undergoes an abrupt attenuation in the vicinity of the cell designated CL, corresponding to Clipperton Atoll, the westernmost member of $B$.

Significantly weaker reactive currents are observed to emanate from the cells labeled HN, HC, and HM, which are associated with the Hawaiian archipelago. The reduced magnitude of these currents is attributed to the presence of a local maximum of the stationary distribution within the Great Pacific Garbage Patch (GPGP) \cite{Moore-eta-al-01}, located east of Hawaii along the pathway of the reactive currents. The computed density of reactive trajectories---quantifying the spatial distribution of time spent by reactive trajectories during transitions from source to target---attains its maximum precisely in this region, as illustrated in right panel of Figure~\ref{fig:stationary}. In contrast, the reactive currents originating from the Line Islands do not encounter a comparable bottleneck en route to the target and are therefore substantially stronger.

\begin{figure}[t!]
    \centering
    \includegraphics[width=.49\linewidth]{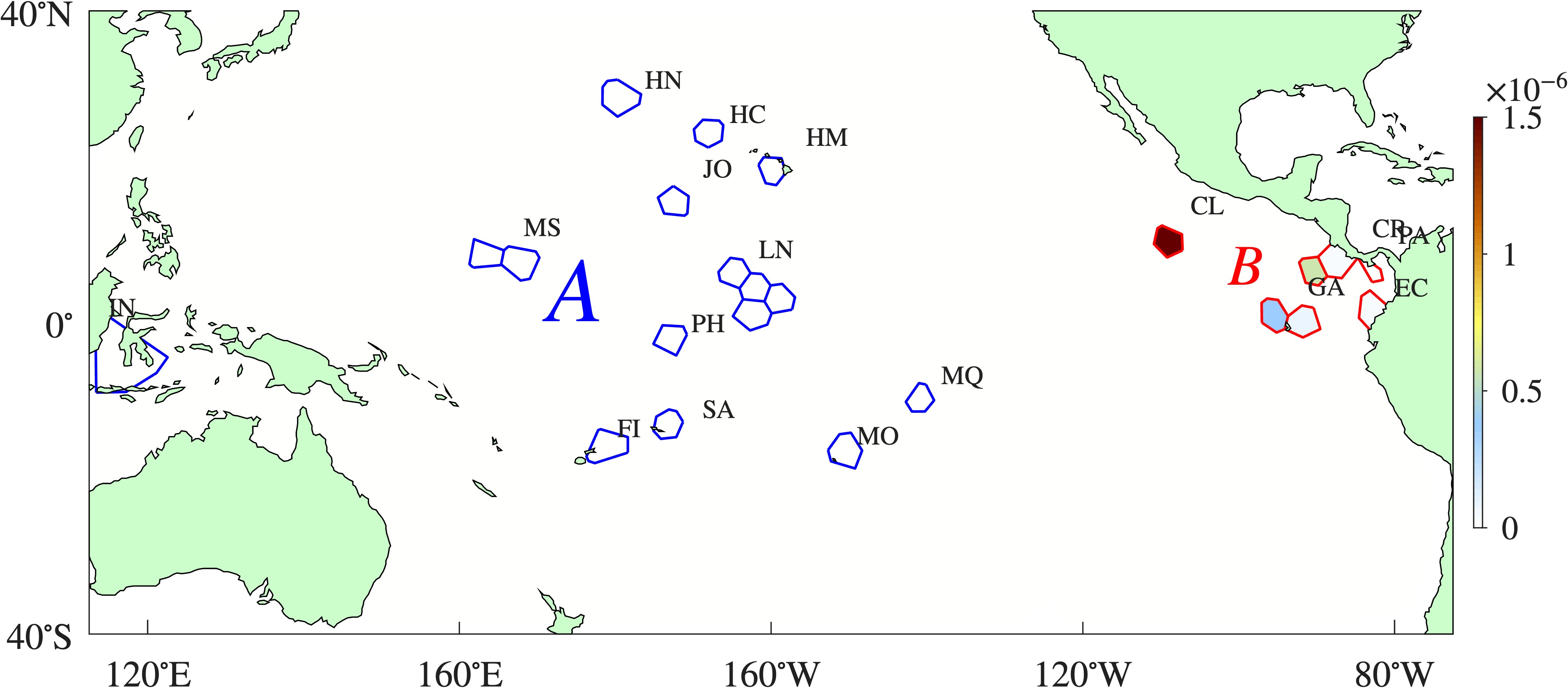},\,
    \includegraphics[width=.49\linewidth]{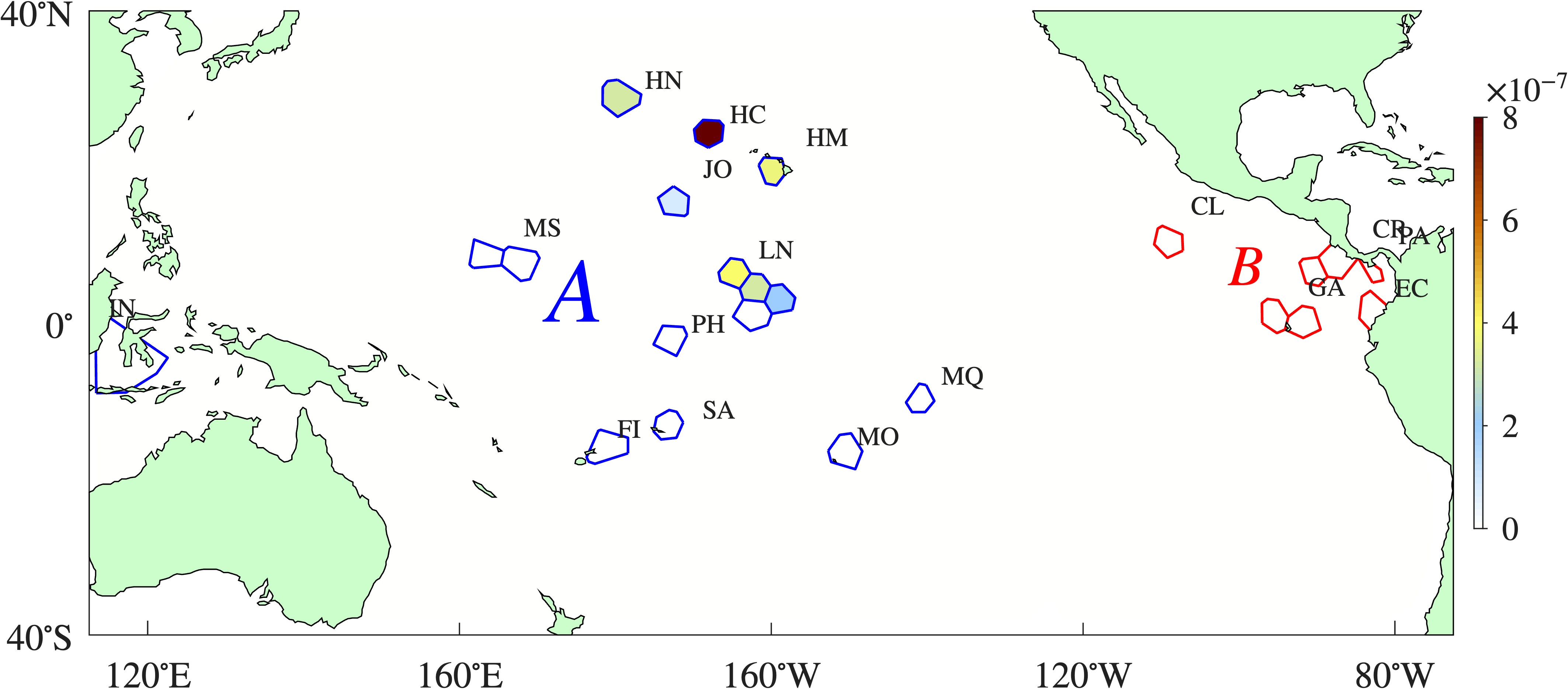}
    \caption{Arrival (left) or departure (right) rate, or probability per time step, of a reactive trajectory reaching each state $i$ that constitutes $A$ or leaving each state $i$ that constitutes $b$, computed by summing the reactive flux entering that particular state $i$ or exiting that particular state $i$.}
    \label{fig:k}
\end{figure}

The spatial distribution within the target set $B$ at which reactive currents originating from the source set $A$ arrive can be characterized by computing the arrival rate into each $i \in B$, which we have denoted $k^{i\in B\leftarrow}$. This quantity is illustrated in the left panel of Figure~\ref{fig:k}. The arrival rate $k^{i\in B\leftarrow}$ attains its maximum in the cell CL, spatially aligned with Clipperton Atoll.  Analogously, one can determine the regions within $A$ from which the largest portion of reactive current emanates by computing $k^{i\in A\to}$, that is, the departure rate from each $i \in A$. A priori, one would expect $k^{i\in A\to}$ to attain its maximum in the set of cells LN. However, as shown in the right panel of Figure~\ref{fig:k}, although $k^{i\in A\to}$ does exhibit a local maximum in one cell of LN, its global maximum occurs in the cell HC, located within the Hawaiian archipelago.  This observation is nevertheless consistent with the interpretation that the net strength of the reactive current directly exiting HL is substantially weaker than that exiting LN. The reactive currents originating in HL tend to meander within the GPGP region en route to $B$, with some trajectories diverting to visit LN before subsequently intensifying and eventually reaching CL, the Clipperton Atoll.

\begin{figure}[t!]
    \centering
    \includegraphics[width=\linewidth]{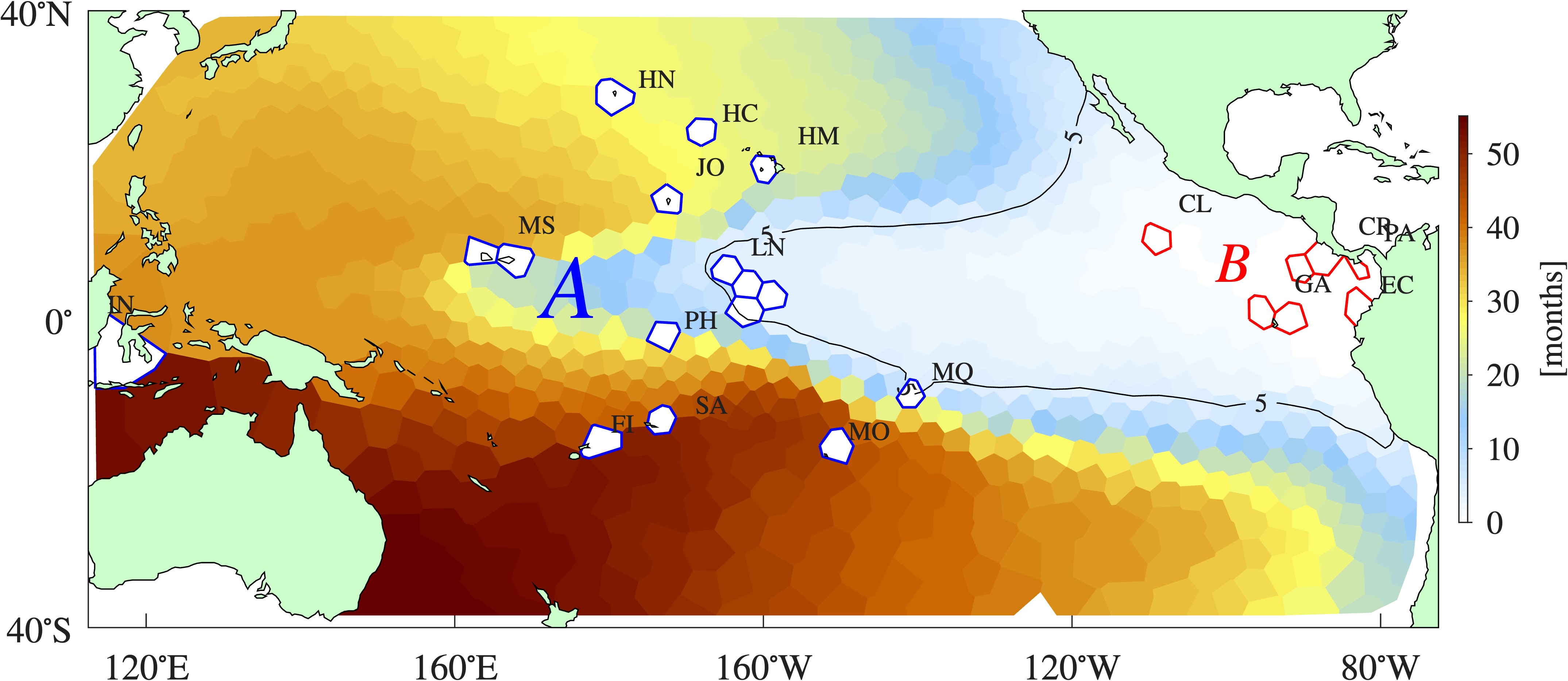}
    \caption{Remaining duration of reactive trajectories, or the expected first entrance time to $B$ from any state $i$ after leaving $A$ on a direct route to $B$, with the 5-month level set highlighted.}
    \label{fig:t}
\end{figure}

Setting aside biological interpretations for the reef-building coral \emph{P.\ lobata}, the results suggest possible connectivity between the western and eastern flanks of the EPB. This is consistent with the idea of weak impermeability of the EPB \cite{Baums-etal-12} based on physical (i.e., circulation of the Pacific Ocean) rather than biological grounds, and suggests the need to refine the definition of the EPB on such grounds. Within the TPT framework, such a refinement can be informed by the remaining duration of reactive trajectories from $A$ to $B$, denoted by $t^{iB}$. This quantity is illustrated in Figure~\ref{fig:t}. In that figure, the black curve represents the 5‑month level set of \(t^{iB}\). This value can be interpreted as a relatively sharp boundary that separates predominantly longer remaining durations on its western side from predominantly shorter durations on its eastern side. The function $t^{iB}$ is approximately piecewise linear, with distinct changes in slope occurring at about 5 and 37 months. We do not display the 37‑month isoline in the figure, because this temporal range is not relevant from a biological point of view, as discussed below.

As expected, the closer a state $i \in D$ is to $B \subset D$, the shorter the remaining duration $t^{iB}$ becomes. The connectivity between $\text{LN} \subset A$ and $B$, in particular with $\text{CL} \in B$, inferred from the TPT analysis, is consistent with the partition implied by $t^{iB}$, with LN lying approximately east of the inferred boundary. Accordingly, a TPT-based definition of the EPB can be formulated as the region characterized by a rapid change in $t^{iB}$. Needless to say, all states of a Markov chain under the ergodicity assumption, as we have assumed to apply the TPT framework, communicate. However, as TPT demonstrates, this communication can be more efficient between certain states, specifically along transition paths connecting source and target that minimize detours. In this sense, connectivity can be characterized as rapid where the $t^{iB}$ statistic is small, as it further suggests for transition paths between the LN cells (Line Islands) and the CL cell (Clipperton Atoll) across the EPB, that is, within a region of rapid connectivity according to the refined EPB definition.

\begin{figure}[t!]
    \centering
    \includegraphics[width=.49\linewidth]{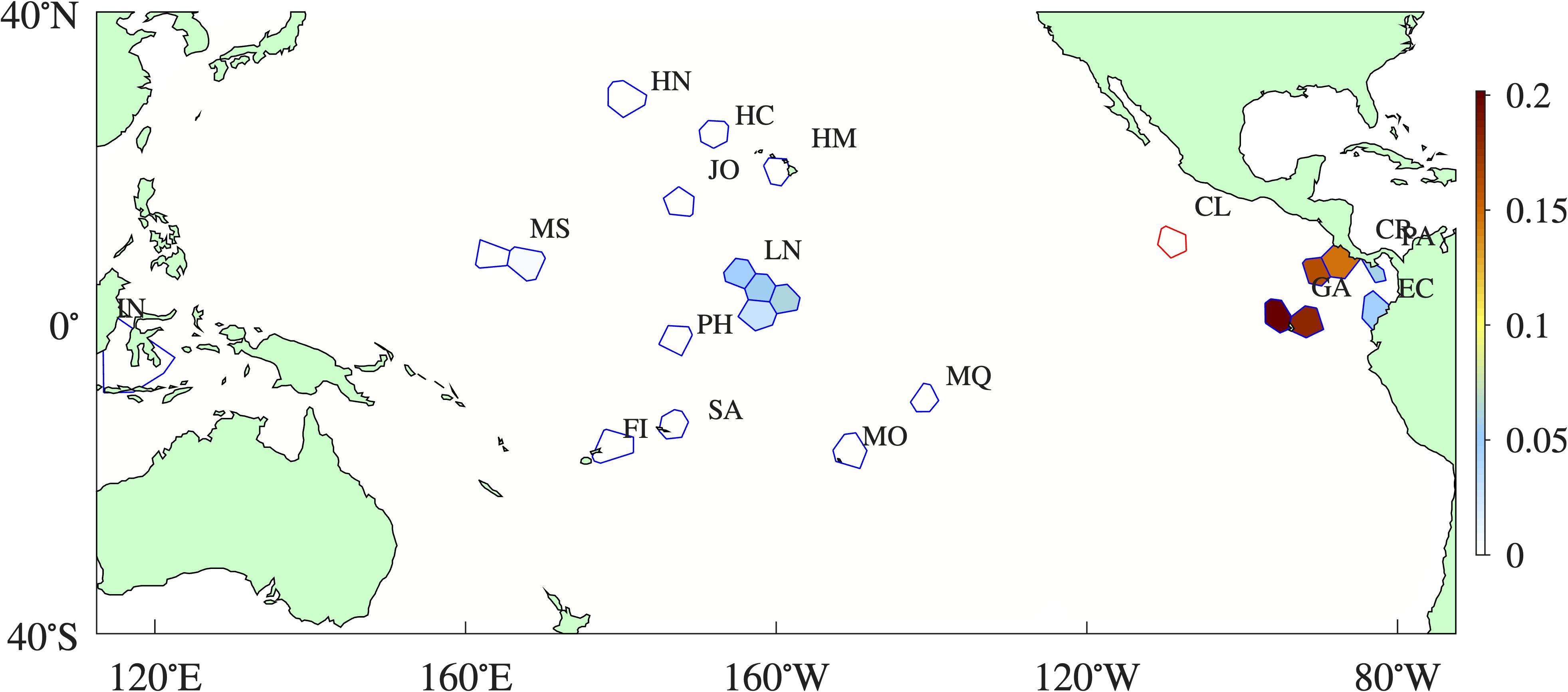}\,
    \includegraphics[width=.49\linewidth]{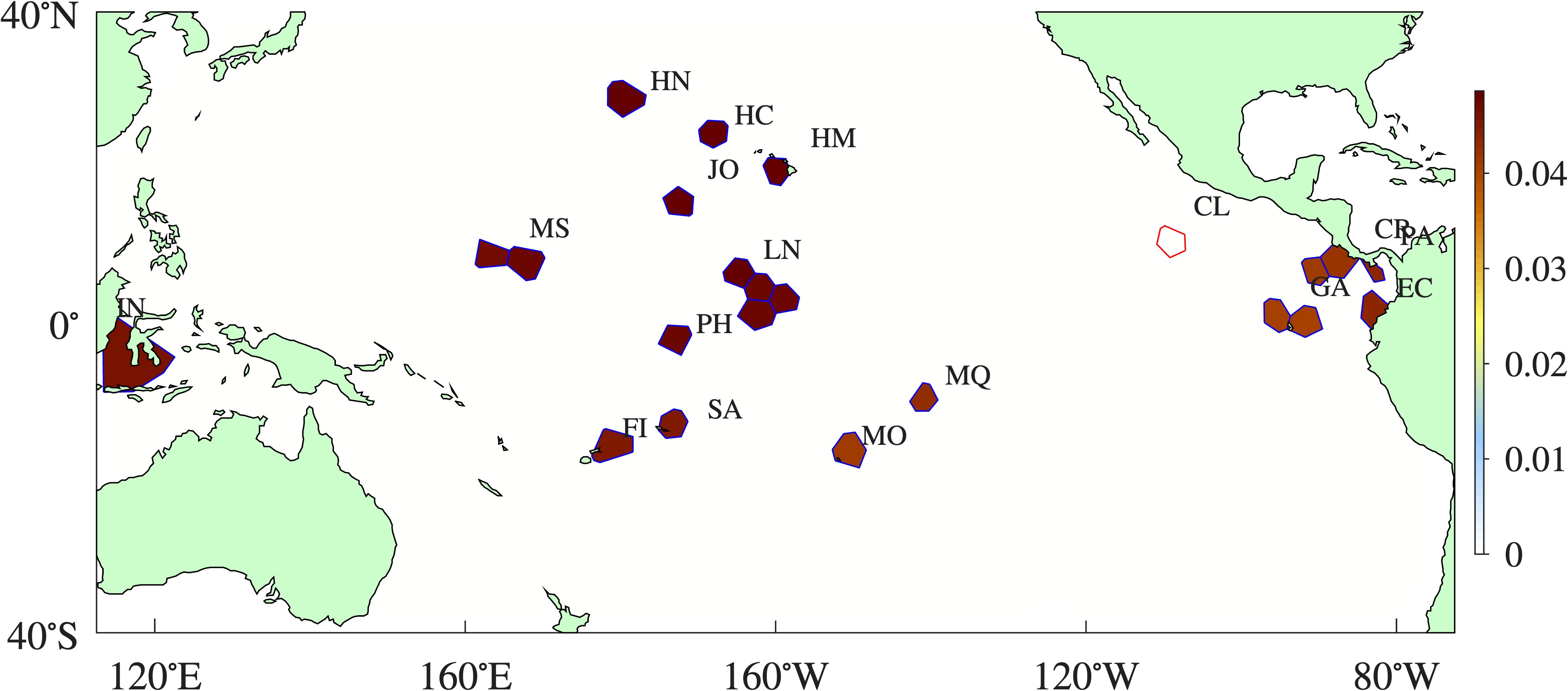}\\\medskip
    \includegraphics[width=.49\linewidth]{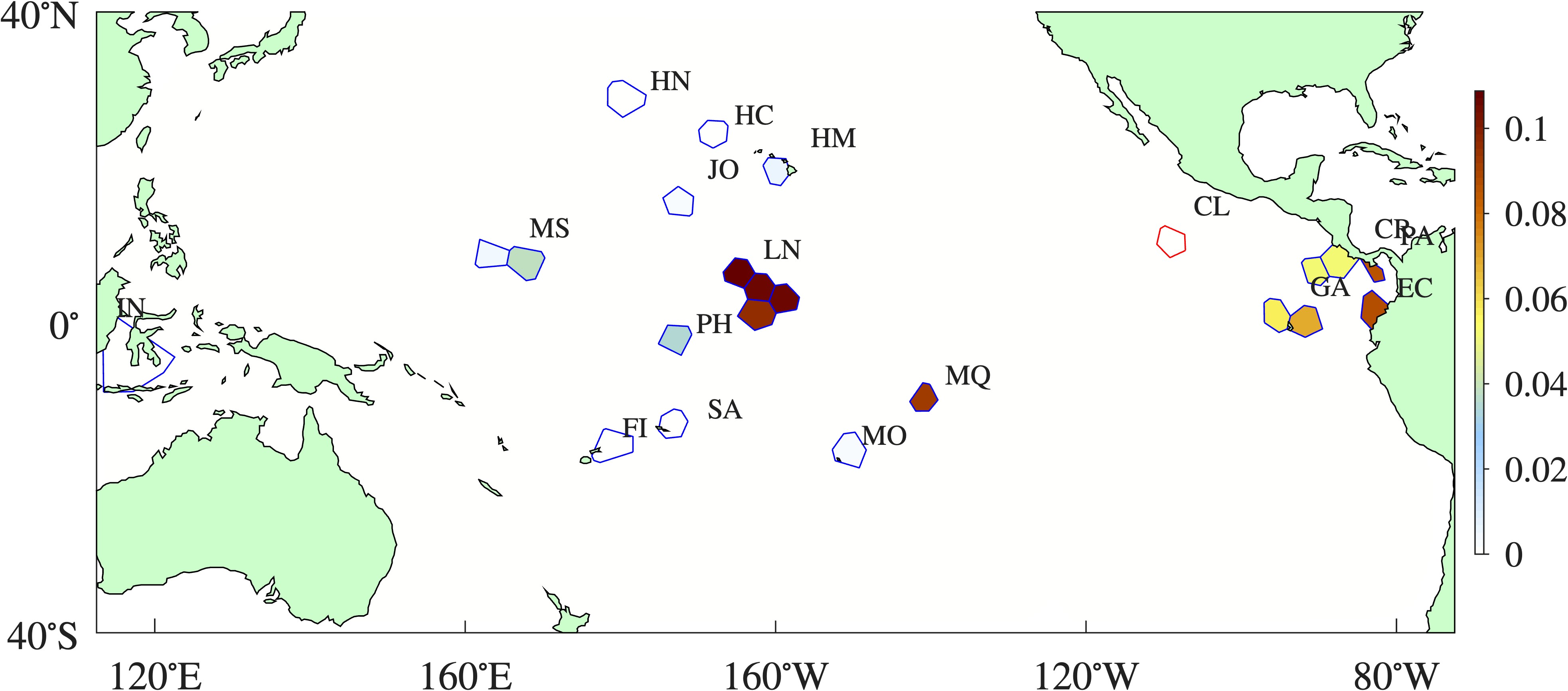}\,
    \includegraphics[width=.49\linewidth]{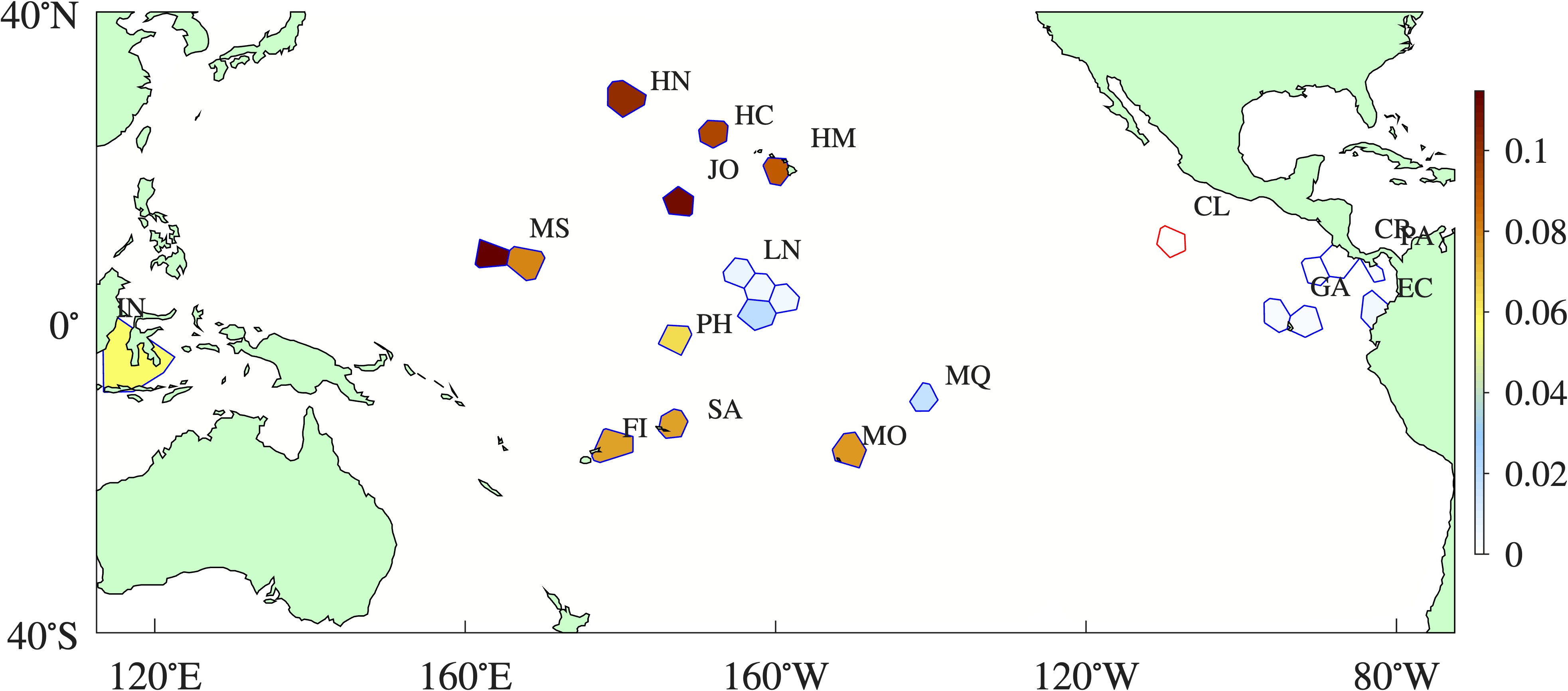}
    \caption{(top panels) Posterior distribution of larval source locations over the set denoted by $\overline{\text{CL}}$, given by the disjoint union of partition cells that intersect each indicated Pacific ``island,'' conditional on the observation that larvae arrive in the partition cell intersecting Clipperton Atoll, labeled CL, 2.5 (left) and 7.5 (right) months after departing from $\overline{\text{CL}}$. (bottom panels) As in the top panels, but using a transition matrix that accounts for direct transitions from $\overline{\text{CL}}$ to CL.}
    \label{fig:bayes}
\end{figure}

We next further examine the distinctive role of the CL cell---specifically, the Clipperton Atoll---as a terminal endpoint for trajectories. To this end, we focus on trajectories that originate in any other cell within the union of $A$ and $B$, excluding CL itself, i.e., $\overline{\text{CL}} := (A \cup B)\setminus\text{CL}$.  The assessment is carried out using Bayesian inference by computing the posterior probability $p(\overline{\text{cl}} \mid t^\text{CL})$ of the origin in $\overline{\text{cl}} \in \overline{\text{CL}}$ of trajectories reaching CL after being observed at time $t^\text{CL}$. The Bayesian inference is first performed while ignoring the specific way in which $\overline{\text{CL}}$ and CL may be connected; that is, direct connectivity as in TPT is not assumed. The corresponding results are displayed in the upper panels of Figure~\ref{fig:bayes}, under the assumption that trajectories are observed to arrive in CL 2.5 months (left panel) and 7.5 months (right panel) after departing from $\overline{\text{CL}}$. The observation times in CL, $t^\text{CL}$, are chosen to be, respectively, shorter and longer than the level set $\{t^{iB} = 5\,\text{months}\}$, which has been used to refine the definition of the EPB (Figure~\ref{fig:t}). As expected, for shorter observation times, such as $t^\text{CL} = 2.5$ months, the posterior probability $p(\overline{\text{cl}} \mid t^\text{CL})$ attains its largest values in grid cells that are geographically closest to CL, such as the GA cells in the Galapagos region. For longer observation times, such as $t^\text{CL} = 7.5$ months, $p(\overline{\text{cl}} \mid t^\text{CL})$ becomes approximately uniform over the complement $\overline{\text{CL}}$, consistent with the ergodicity of the underlying Markov chain. 

A more informative characterization is obtained by restricting attention to direct (i.e., minimally detouring) transitions from $\overline{\text{CL}}$ to $\text{CL}$. This analysis employs $P^+$, as defined in \eqref{eq:P+}, that is, the forward–reactive variant of the transition matrix $P$, to perform the Bayesian inference. The $P^+$-based results, with $A = \overline{\text{CL}}$ and $B = \text{CL}$, are presented in the lower panels of Figure~\ref{fig:bayes}. Under this constraint, for $t^{\text{CL}} = 2.5$ months, $p(\overline{\text{cl}} \mid t^\text{CL})$ attains its maximum in the cells labeled LN, corresponding to the Line Islands. This finding is consistent with the imposed constraint and with the spatial distribution of the remaining duration of reactive trajectories, $t^{iB}$, depicted in Figure~\ref{fig:t}. For longer observation times, such as $t^{\text{CL}} = 7.5$ months, $p(\overline{\text{cl}} \mid t^{\text{CL}})$ begins to take larger values in those cells of $\overline{\text{CL}}$ that are geographically located west of the $t^{iB}$ isoline of 5 months.

In the following section, we examine the implications of these results---derived from the surface ocean Lagrangian circulation in the Pacific---for the dispersal and population connectivity of \emph{P.\ lobata}.

\section{Discussion}

The Reef-building \emph{P.\ lobata} larvae show substantial variation in their dispersal potential. Because their eggs contain symbiotic algae, larvae can be nourished throughout their time in plankton, which may increase the extent to which they are able to disperse \cite{Baums-etal-12}. While many individuals in an ex situ (i.e., controlled, laboratory) experiment reach settlement competency as early as 2 days post-fertilization  \cite{Bennett-etal-24}, biophysical connectivity models for the Eastern Tropical Pacific  often assume pelagic larval durations (PLDs) on the order of several weeks to about 2 months to represent typical dispersal. Longer maximum PLDs of roughly 120--150 days are commonly implemented to explore rare trans-oceanic connectivity across the Eastern Pacific Barrier \cite{Romero-etal-18, Wood-etal-16}. Even broader PLD limits of more than 200 days have been used in global dispersal simulations to capture the extreme outliers required to maintain large-scale genetic connectivity among coral populations \cite{Wood-etal-14}.

The above characteristics must be explicitly considered when interpreting our results in the context of \emph{P.\ lobata} connectivity across the EPB. Consequently, the remaining duration of reactive trajectories, $t^{iB}$, constitutes a particularly pertinent TPT diagnostic to monitor. Indeed, as shown in Figure~\ref{fig:t}, the 5‑month isoline carries a biological meaning that extends beyond its physical interpretation for \emph{P.\ lobata} connectivity. Only those reactive trajectories linking LN cells, associated with the Line Islands, to the CL cell, representing Clipperton Atoll, can thus be assigned biological relevance. Linkage with other ``islands'' is not biologically realizable, as their $t^{iB}$ values are (much) longer than the 5‑month survival horizon. 

Connectivity between Clipperton Atoll, represented by the CL cell, and the Hawaiian Archipelago, represented by cells HN, HC, and HM---which exhibit a high departure rate $k^{i\in A\to}$ (Figure~\ref{fig:k}, right panel)---can now be considered biologically implausible, because the transit time required for particles to reach CL substantially exceeds the pelagic larval duration (survival threshold) of \emph{P.\ lobata}. This result reconciles our circulation-based results for near-surface Pacific flow with the conclusions derived solely from genetic studies in \cite{Baums-etal-12}. As discussed above, the Northern Hemisphere subtropical gyre strongly influences the very long $t^{iB}$ values associated with reactive trajectories originating in Hawaii (HN, HC, and HM cells), owing to the role of the gyre center as an attractor for Lagrangian trajectories, driven by Ekman convergence.

\subsection{Seasonal Variability}

The reactive current connecting the Line Islands (LN) and Clipperton Atoll (CL) should evidently modulated by the NECC variability. Driven by the seasonal meridional migration of the ITCZ and westward-propagating Rossby waves, the NECC typically intensifies during boreal summer--fall and weakens or even vanishes during winter--spring \cite{Philander-etal-87, Hsin-Qiu-12a}. This motivates an investigation of the sensitivity of the TPT results to seasonal changes.

To perform this sensitivity analysis, we construct two Markov chain models: one based on GDP drifter trajectories that overlap with summer--fall months, and another based on trajectories overlapping with winter--spring months. These seasonal constraints reduce the number of available GDP drifter trajectory pairs connecting times $t$ and $t+5$ days, for any $t$ in the period 1979--2025, from approximately $2\times 10^6$ to about $1.5\times 10^6$ for both the summer--fall and winter--spring subsets. The number of cells in the domain partition is fixed at $10^3$. However, this does not lead to identical Voronoi tessellations, as these are determined by the spatial distribution of the trajectory pairs.

\begin{figure}[t!]
    \centering
    \includegraphics[width=.75\linewidth]{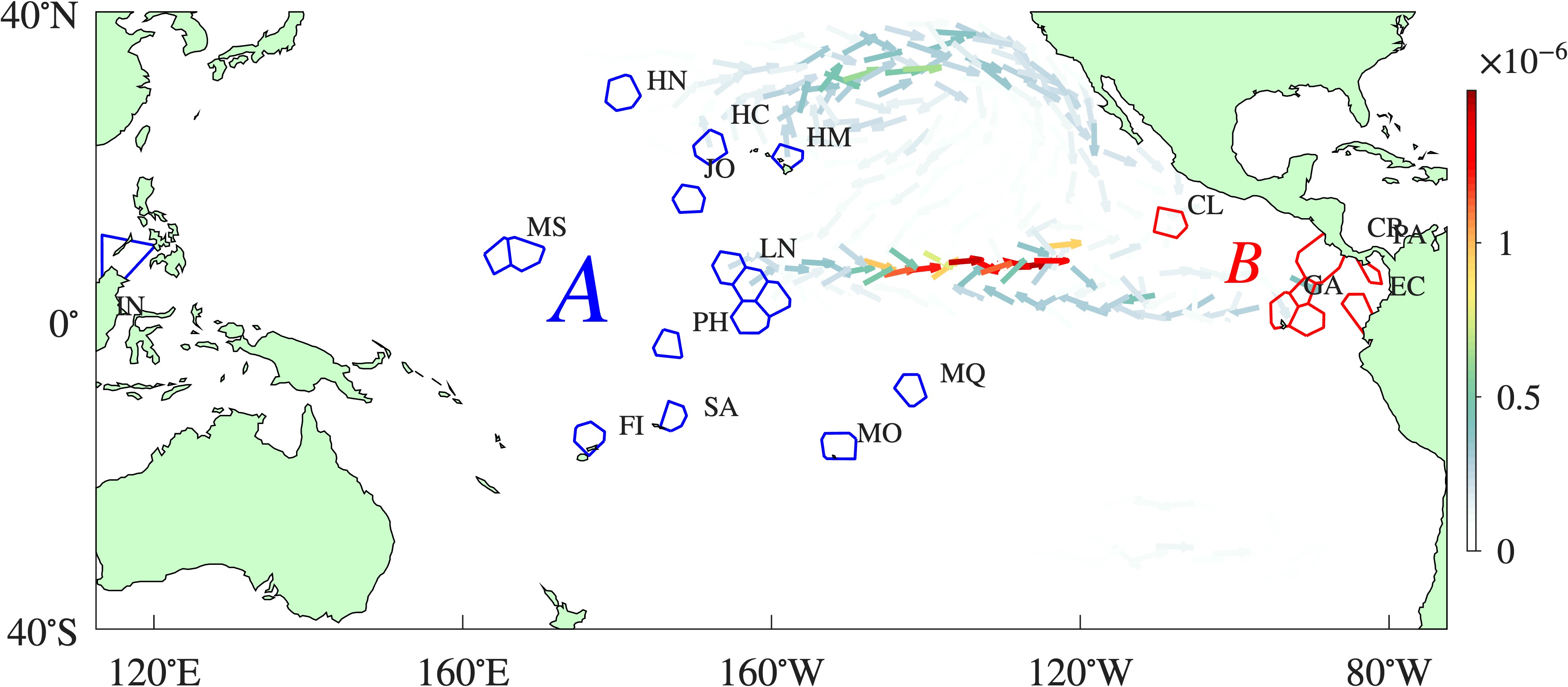}\\\medskip
    \includegraphics[width=.75\linewidth]{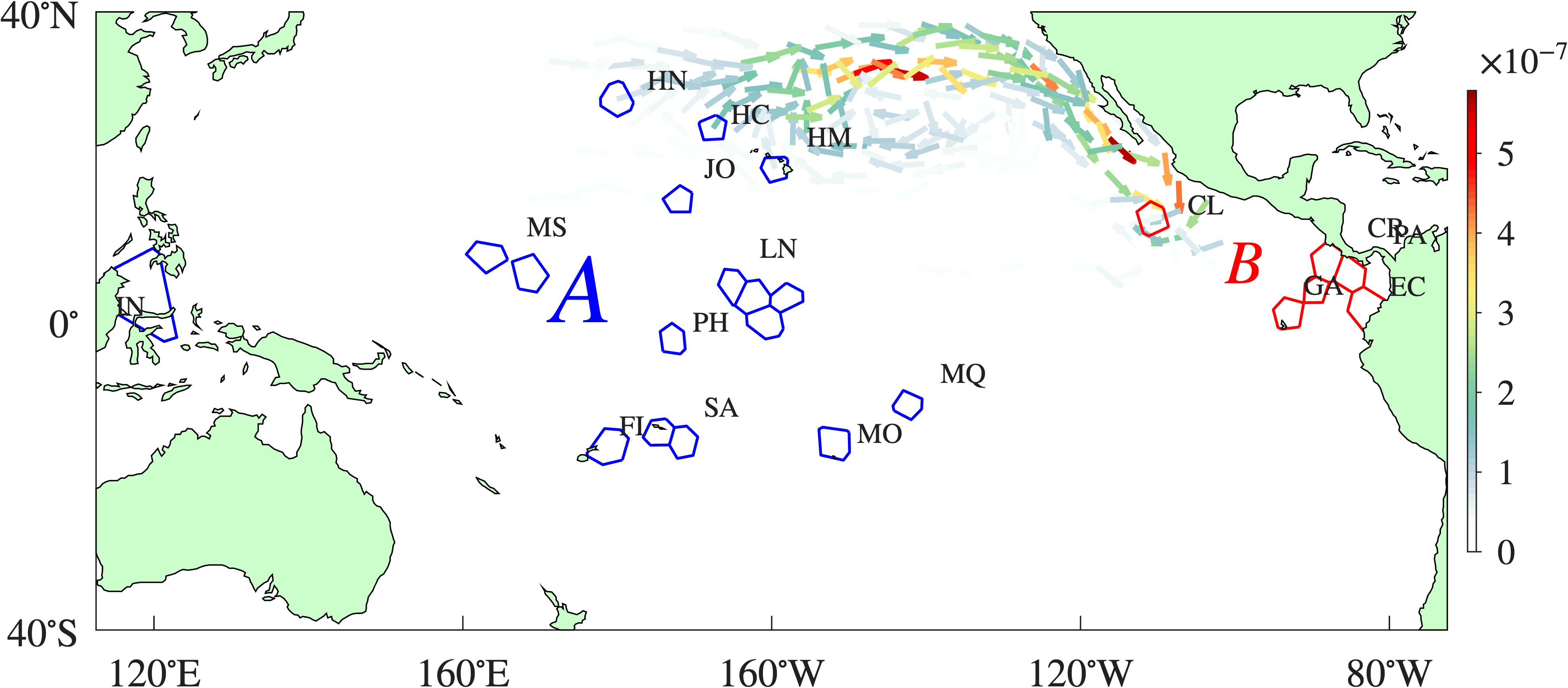}
    \caption{As in Figure~\ref{fig:currents}, but with the Markov chain constructed using trajectories over summer--fall (top) and winter--spring (bottom) periods.}
    \label{fig:season}
\end{figure}

As above, the TPT analysis is performed by defining the source set $A$ and the target set $B$ as the disjoint unions of the cells associated with the ``islands'' located west and east of the EPB, respectively. The resulting reactive currents are presented in Figure~\ref{fig:season}, with those corresponding to summer--fall (winter--spring) displayed in the upper (lower) panel. Note that the scale used here differs from that in Figure~\ref{fig:currents}, which must be borne in mind when making a comparison. A reactive current linking the Line Islands (LN cells) and Clipperton Atoll (the CL cell) is clearly present in summer--fall, but it disappears in winter--spring. This seasonal contrast is consistent with the variability of the NECC described above. The current that develops during El Ni\~no phases is of comparable magnitude to that obtained when only neutral periods are considered. In contrast, the reactive current during La Ni\~na phases that links the Hawaiian archipelago (cells NH, HC, and HM) with Clipperton Atoll is at least one order of magnitude weaker than that observed either during El Ni\~no phases or when both El Ni\~no and La Ni\~na periods are excluded from the analysis.

\subsection{Interannual Variability}

The NECC generally strengthens and shifts northward during El Ni\~no events, while it weakens and moves southward during La Ni\~na events \cite{Wyrtki-73b, Hsin-Qiu-12b}. This interannual variability in both the intensity and location of the NECC prompts a further investigation of how sensitive the TPT results are to these ENSO-related changes.

In the construction of the Markov chains thus far, we have excluded all trajectories that overlap with intervals classified as El Ni\~no or La Ni\~na according to the Oceanic Niño Index (ONI) \cite{ONI-26}. Without distinguishing by season as in the previous analysis, we now repeat the TPT analysis while restricting the dataset to trajectories that overlap with periods characterized by $\text{ONI} > +0.5^\circ\text{C}$ (El Ni\~no) and $\text{ONI} < -0.5^\circ\text{C}$ (La Ni\~na), which are analyzed separately. These constraints reduce the number of available GDP drifter trajectory pairs to approximately $0.5\times 10^6$ for both El Ni\~no and La Ni\~na conditions. As before, the number of cells in the spatial partition of the domain is fixed at $10^3$, while noting that the resulting Voronoi tessellations differ between the two cases due to the distinct spatial distributions of the trajectory pairs.

\begin{figure}[t!]
    \centering
    \includegraphics[width=.49\linewidth]{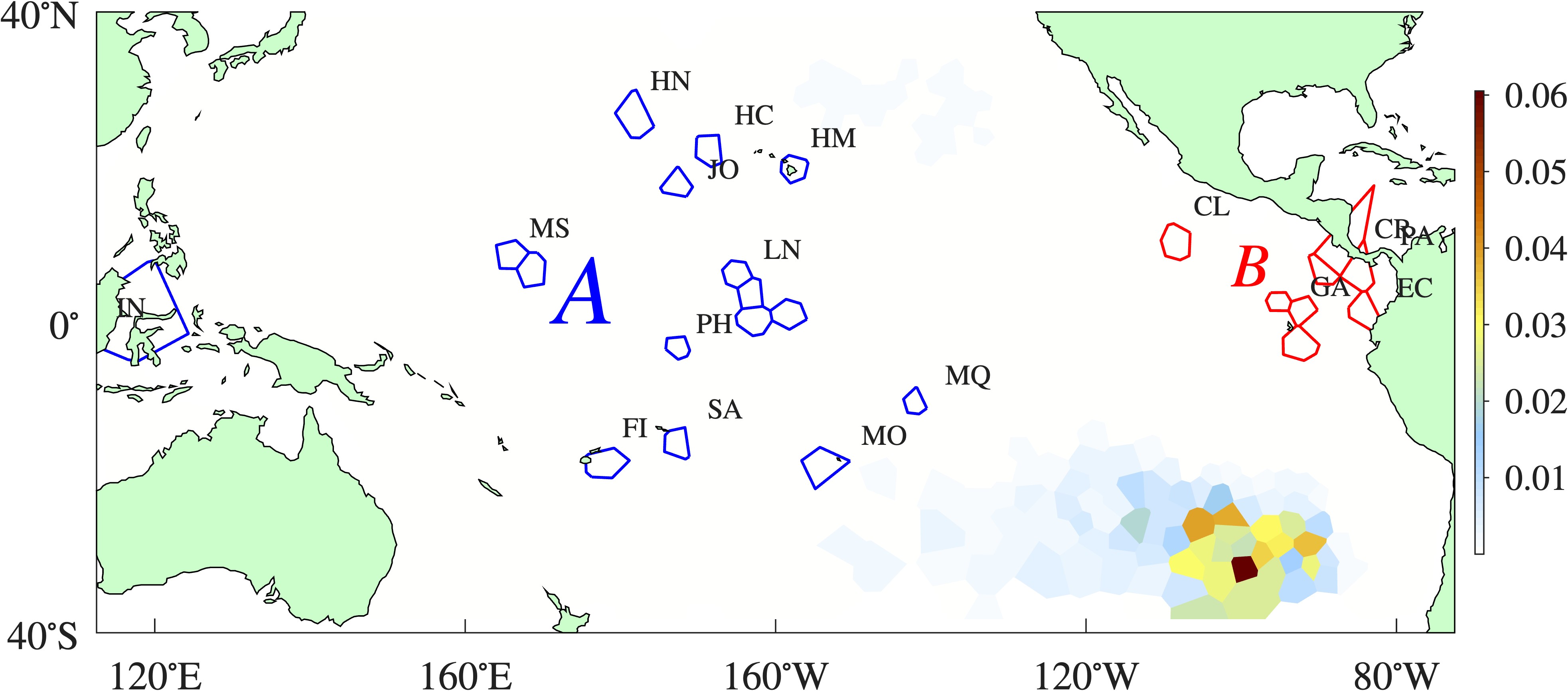}\,
    \includegraphics[width=.49\linewidth]{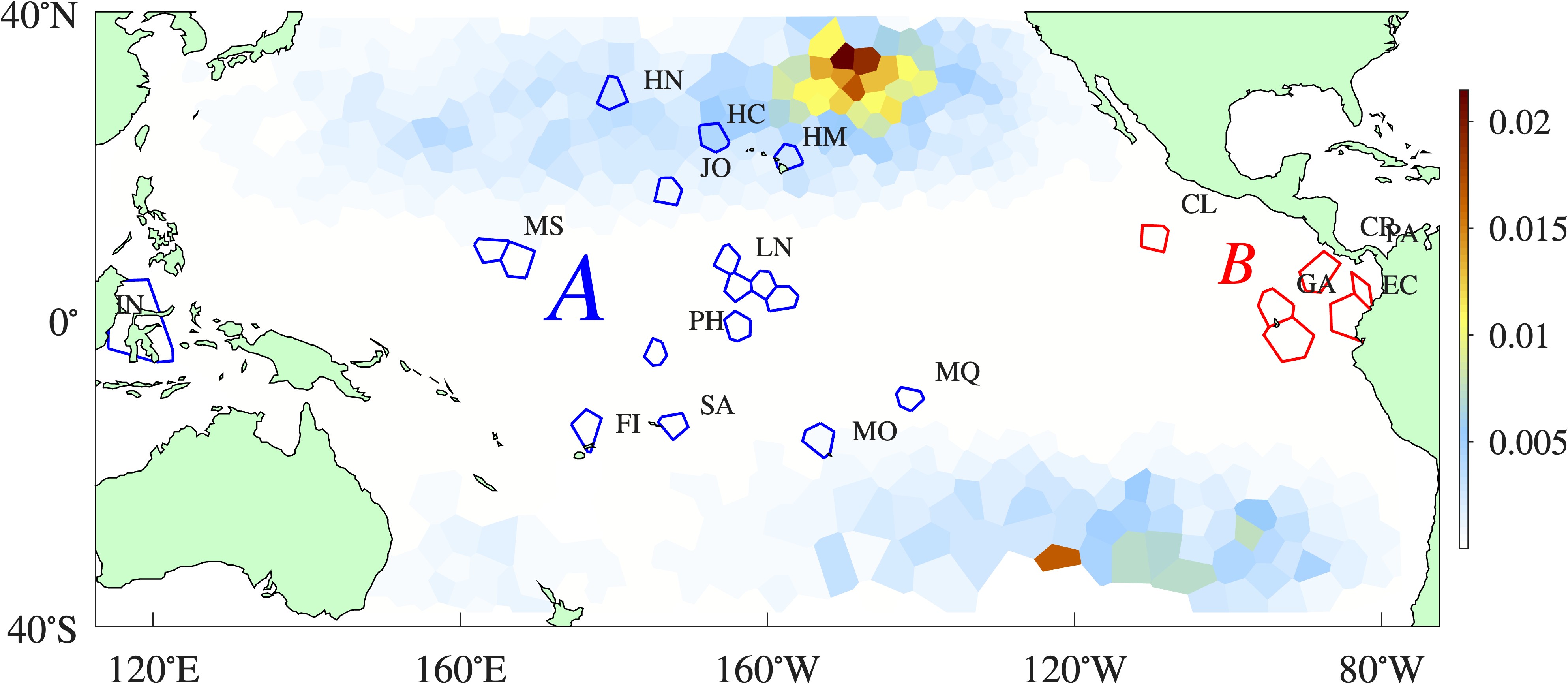}\\
    \medskip
    \includegraphics[width=.49\linewidth]{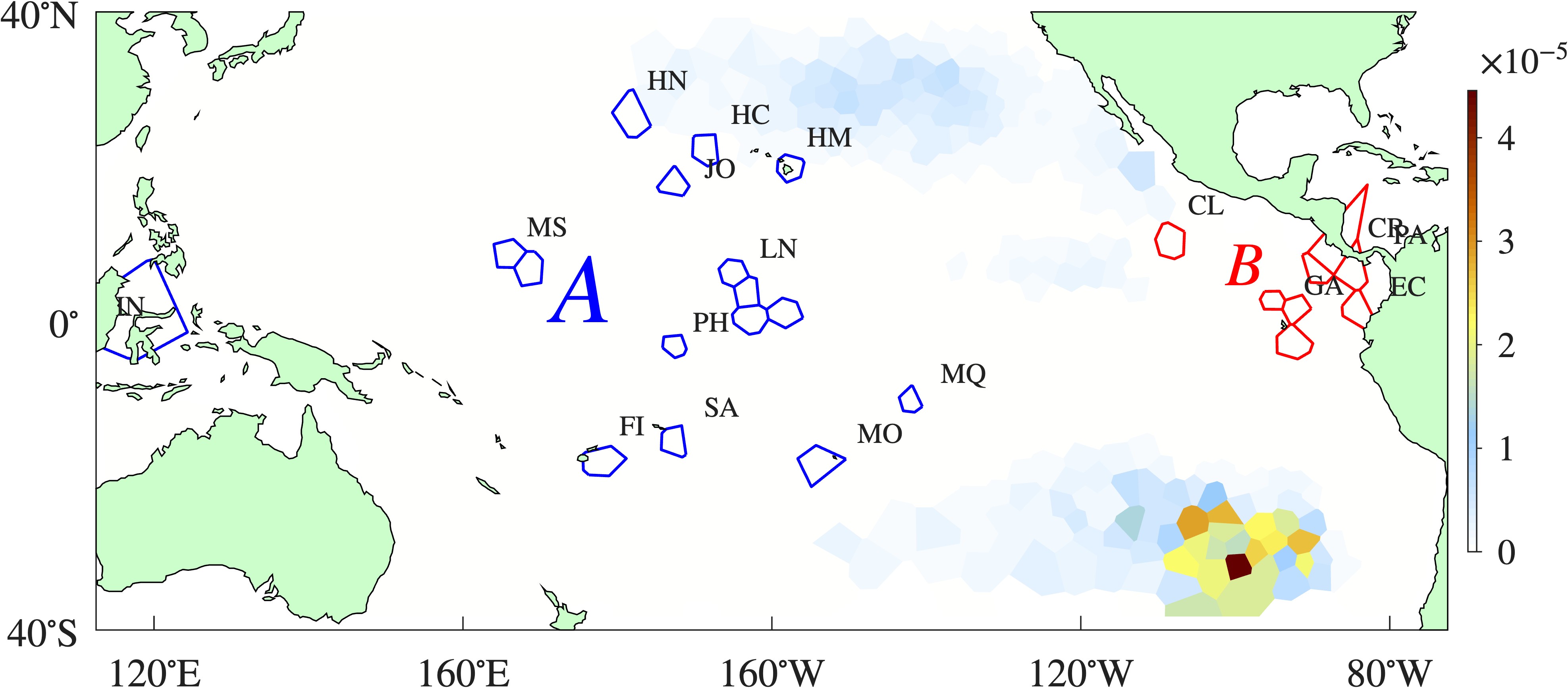}\,
    \includegraphics[width=.49\linewidth]{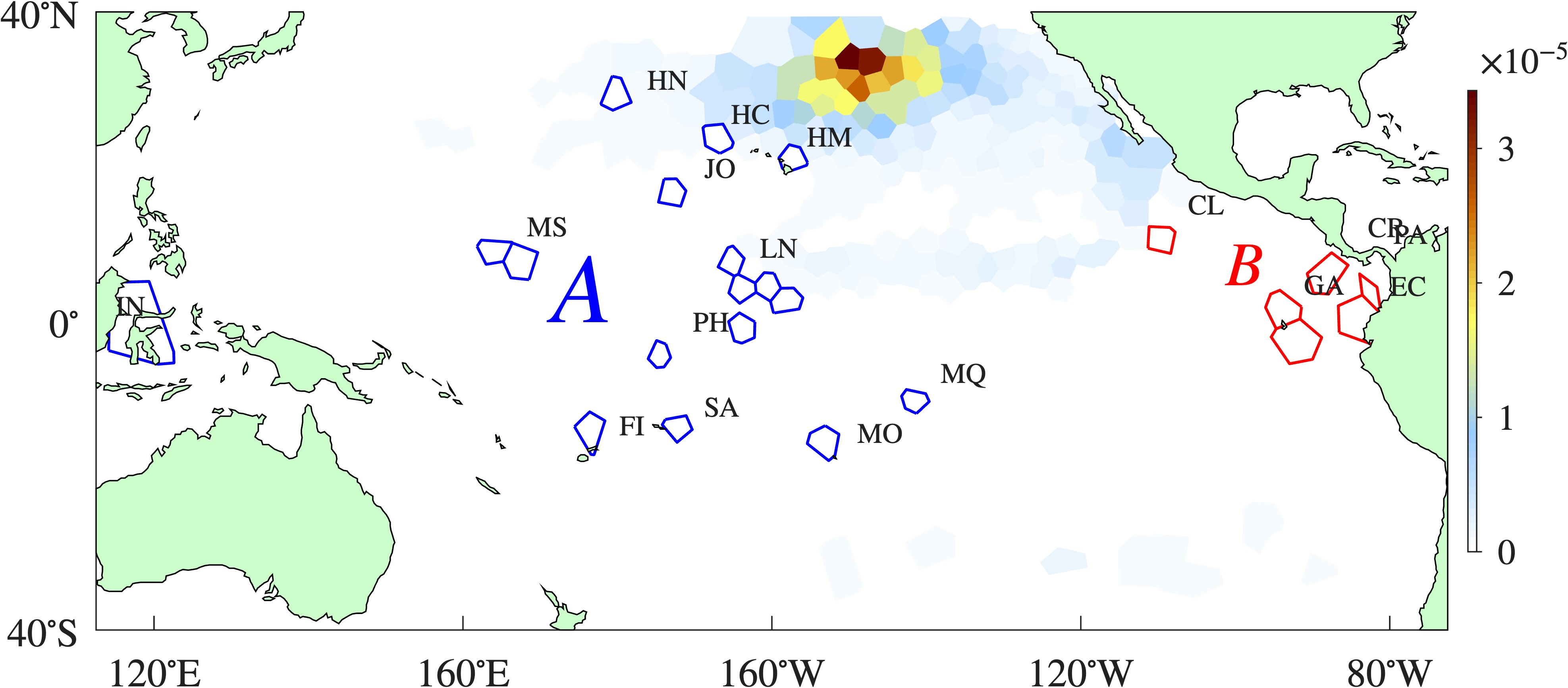}\\
    \medskip
    \includegraphics[width=.49\linewidth]{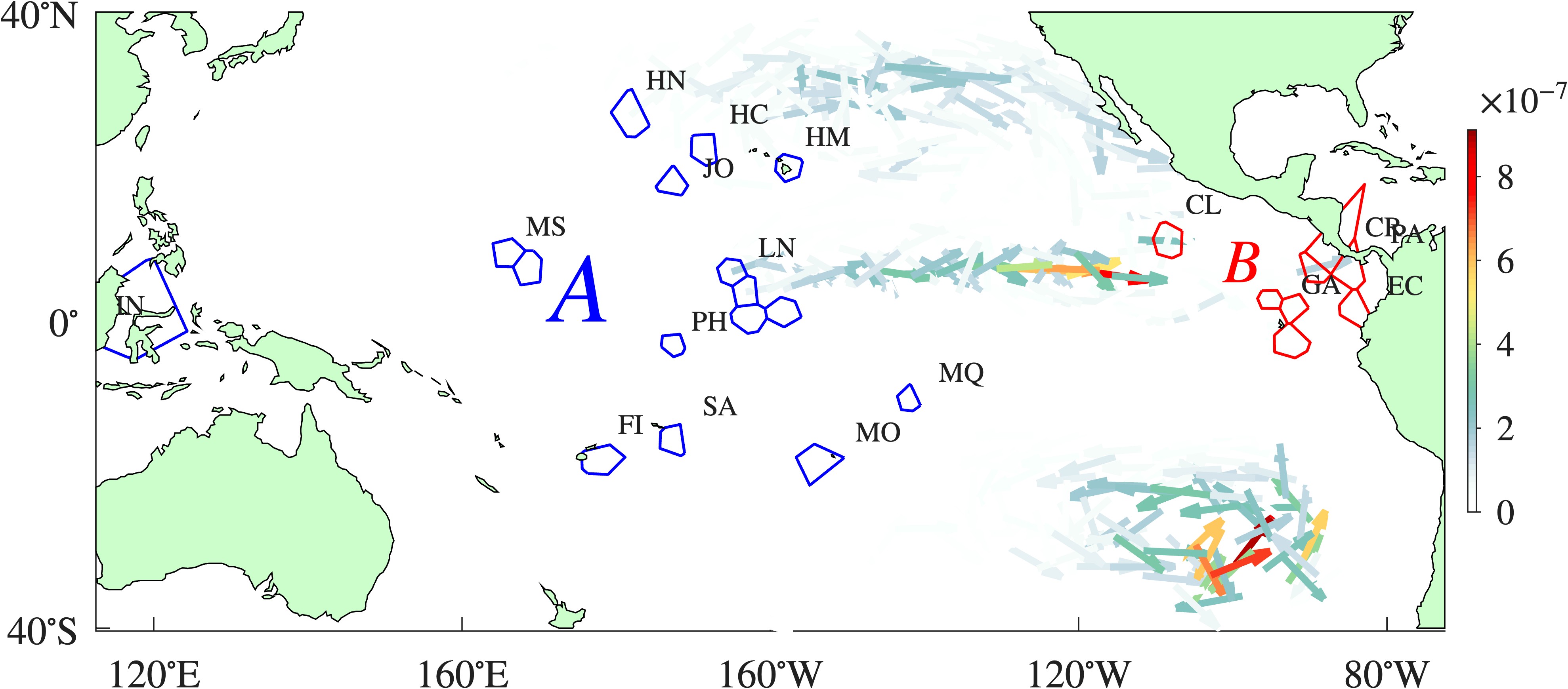}\,
    \includegraphics[width=.49\linewidth]{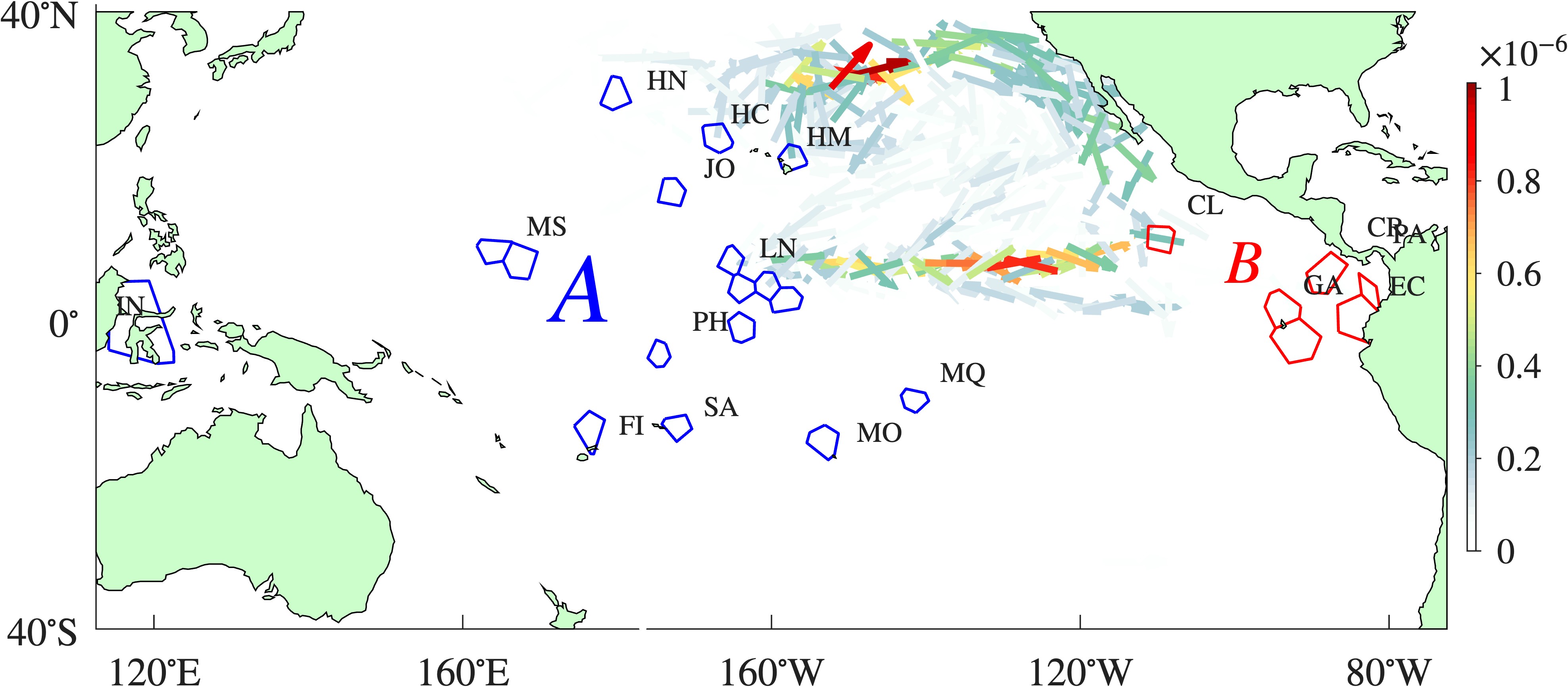}
    \caption{(top panels) As in the left panel of Figure~\ref{fig:stationary}, but with the Markov chain constructed using trajectories over El Ni\~no (left) La Ni\~na (right) periods. (middle panels) As in the right panel Figure~\ref{fig:currents}, but with the Markov chain constructed using trajectories over El Ni\~no (left) La Ni\~na (right) periods.  (bottom panels) As in Figure~\ref{fig:currents}, but with the Markov chain constructed using trajectories over El Ni\~no (left) La Ni\~na (right) periods.}
    \label{fig:nino-nina}
\end{figure}

The left column of Figure~\ref{fig:nino-nina} presents the results obtained under El Ni\~no conditions, while the right column displays those corresponding to La Ni\~na conditions. In addition to the reactive currents (bottom panels), we show the corresponding stationary distribution $\boldsymbol\pi$ (top panels) and the density of reactive trajectories $\boldsymbol\mu$ (middle panels), as these statistics are seen depend on ENSO phase (they are effectively independent of the seasonal cycle and thus were not discussed above). Note that the scales used in this figure differ from those in the analogous figures shown earlier. This distinction is critical for interpreting the reactive current behavior.

Under El Ni\~no conditions, the stationary distribution $\boldsymbol\pi$ attains its maximum only in the subtropical gyre of the Southern Hemisphere. This is consistent with the fact that El Ni\~no is characterized by weakened or reversed equatorial trade winds, which allow warm surface waters from the western tropical Pacific to propagate eastward toward the eastern Pacific and the American continent.

Under La Ni\~na conditions, which may be interpreted as an amplification of neutral conditions in terms of flow directionality, $\boldsymbol\pi$ exhibits maxima in the subtropical gyre centers of both hemispheres. The density of reactive trajectories $\boldsymbol\mu$ attains its maximum in the Northern (Southern) Hemisphere subtropical gyre during El Ni\~no (La Ni\~na) conditions, indicating the regions where reactive pathways tend to experience bottlenecks. This, in turn, shapes the structure of the reactive currents, which are most pronounced along the route from the Line Islands (cells labeled LN) to Clipperton Atoll (cell labeled CL). The magnitude of this reactive current is, however, weaker than the corresponding current under neutral conditions.

Another difference to be noted concerns the meridional deflection of the reactive pathways between LN and CL: during El Ni\~no conditions they exhibit a slight northward curvature, whereas during La Ni\~na they bend slightly southward. This behavior appears to be in agreement with with the documented meridional displacement of the NECC. However, the anticipated strengthening of the NECC during El Ni\~no events and its weakening during La Ni\~na events, as well as the associated impacts on \emph{P.\ lobata} connectivity across the EPB inferred from numerical modeling studies \cite{Wood-etal-16}, are not clearly reflected in the TPT diagnostics. This discrepancy may possibly result from the limited number of trajectory pairs available to construct Markov chains in the El Ni\~no and La Ni\~na subsets, which constrains the statistical robustness of the TPT analysis. We emphasize, however, the observational nature of our assessment.

\section{Conclusions}

The Markov chain analysis of satellite-tracked, i.e., observed, surface drifter trajectories presented in this study reveals that the Eastern Pacific Barrier (EPB) functions as only a weakly impermeable boundary to dispersal by surface currents, consistent with the observed weak genetic connectivity of \emph{Porites lobata} across this region. In particular, we identify efficient, short-duration reactive pathways connecting the Line Islands to Clipperton Atoll that remain within the estimated larval survival window for this coral species, and we determine that the Line Islands represent a local maximum in the posterior probability of successful transport to Clipperton at approximately 2.5 months. Collectively, these findings reconcile oceanographic transport pathways with observed genetic patterns and suggest that the EPB should not be conceptualized as a static or absolute barrier. Instead, it should be redefined in terms of the residual duration and structural characteristics of reactive trajectories that facilitate rare but evolutionarily significant cross-barrier gene flow. Moreover, the connectivity between the Line Islands and Clipperton is primarily modulated by the seasonal variability of the North Equatorial Countercurrent, rather than by the phase of the El Ni\~no--Southern Oscillation (ENSO). The role of Clipperton Atoll as a sink of dispersal trajectories may have implications for conservation in the Clarion--Clipperton Zone (CCZ), the primary region targeted for polymetallic nodule deep-sea mining \cite{Lodege-etal-14}, as sink regions may enhance recolonization through the accumulation of larvae transported by rare long-distance dispersal pathways.

\section*{Acknowledgments}

Ramona Joss participated in undergraduate summer research on exploratory calculations and training in the application of transition path theory to data analysis.

\section*{Supplementary Material}

Supplementary material available online.

\section*{Funding}

This research and manuscript received no external funding.

\section*{Author Contributions}

MJO: Conceptualization; Formal analysis; Interpretation of the results; Writing---review \& editing.  FJBV: Conceptualization; Formal analysis; Interpretation of the results; Writing---main text; review \& editing. GB: Interpretation of the results; Code development; Writing---review \& editing. CM: Formal analysis; Interpretation of the results; Writing---review \& editing. 

\section*{Preprints}

A preprint of this article is available at https://doi.org/10.XXXXX/arXiv.XXX.XXXXX.

\section*{Data, Materials, and Software Availability}

The data from the NOAA Global Drifter Program (GDP) used in this article are available from \href{http://www.aoml.noaa.gov/phod/dac/}{http://www.aoml.noaa.gov/phod/dac/}. Seasonal Oceanic Ni\~no Index (ONI) timeseries employed is available from NOAA Climate Prediction Center (CPC) at \href{https://www.cpc.ncep.noaa.gov/data/indices/oni.ascii.txt}{https://www.cpc.ncep.noaa.gov/data/indices/oni.ascii.txt}. Transition matrix computations were performed using the Julia package \texttt{UlamMethod.jl}, distributed from \href{https://github.com/70Gage70/UlamMethod.jl}{https://github.com/70Gage70/UlamMethod.jl}.

\appendix
\renewcommand{\thesection}{\Alph{section}}
\renewcommand{\theequation}{\thesection.\arabic{equation}}

\setcounter{secnumdepth}{1}
\setcounter{equation}{0}
\section{Preliminaries}

Let $S = \{1,2,\dotsc,N\}$ represent a finite state space.  Let in addition $X_n$ represent, at discrete time $n$, an $S$-valued random variable over an implicitly given probability space, with probability measure given by $\mathbb P$.  Physically, $X_n = i \in S$ means that position of a drifter (representing a planula) happens to be found in box $C_i$ of the partition of the domain $D$ at time $n\Delta t$.

Consider a discrete-time Markov chain $\{X_n\}_{n \in \mathbb Z}$ with row-stochastic transition probability matrix $P = (P_{ij})_{i,j\in S}$. As stated in the main text, and repeated here for completeness, we assume that the Markov chain is ergodic, i.e., the matrix $P$ is both irreducible and aperiodic, and that it is time-homogeneous, meaning that $P$ does not depend on $n$. Under these assumptions, there exists a unique stationary distribution $\boldsymbol\pi$ which is invariant under the transition kernel, $\boldsymbol\pi = \boldsymbol\pi P$, and which is also the limiting distribution, $\boldsymbol\pi = \lim_{k\to\infty} \mathbf f P^k$ for any initial probability vector $\mathbf f$. We choose the initial state distribution as $X_0 \sim \boldsymbol\pi$, so that the Markov chain is strictly stationary; in particular, we have $X_n \sim \boldsymbol\pi P^n = \boldsymbol\pi$ for all $n \in \mathbb Z$.

\setcounter{equation}{0}
\section{Transition path theory}

Here we present the central results of transition path theory (TPT).  Additional details can be found in \citep{E-VandenEijnden-06, VandenEijnden-06, Metzner-etal-09, Helfmann-etal-20, Bonner-etal-23}. 

\subsection{The main objects of TPT}

\paragraph{Reactive process.}

We define the \defn{first entrance time} to a set $U \subset S$ as
\begin{equation}\label{eq:first-passage-forward}
    \tau_U^+(n) := \inf_{k \geq 0}\{k : X_{n+k} \in U\},
\end{equation}
and the \defn{last exit time} from $\mathbb S$ as
\begin{equation}
    \tau_U^-(n) := \inf_{k \ge 0}\{k : X_{n-k} \in U\}.
\end{equation}
In any case, $\inf \varnothing := \infty$. The last exit time is a stopping time with respect to the time-reversed process $\{X_{-n}\}_{n \in \mathbb Z}$, namely, the Markov chain on $S$ with transition probability matrix $P^{-} = (P_{ij}^{-})_{i,j\in S}$ whose entries are given by
\begin{equation}
    P^{-}_{ij} := \frac{\pi_j}{\pi_i} P_{ji}.
\end{equation}
Let $A,B \subset S$ with $A \bigcap B = \varnothing$, such that neither set is reachable from the other in a single step. Adopting the nomenclature conventionally employed in the physical chemistry literature, we classify the process as \defn{forward-reactive}, denoted by $R^+$, or as \defn{backward-reactive}, denoted by $R^-$, depending on the realization of the following events: 
\begin{equation} \label{eq:reactive-R-definition}
    R^\pm(n) := \{\tau_B^\pm(n) < \tau_A^\pm(n)\}.
\end{equation}
Then, the process is \defn{reactive} according to the realization of the event $R$, where
\begin{equation}
    R(n) := \left\{R^-(n) \bigcup R^+(n)\right\}.
\end{equation}
In summary, \emph{a trajectory is reactive if its most recent visit to $A \bigcup B$ was to $A$, it is currently outside of $A \bigcup B$, and its next visit to $A \bigcup B$ will be to $B$.} In the jargon of the physical chemistry literature, $A$ denotes the \emph{reactant} and $B$ the \emph{product} of a reactive process. In the context of fluid transport applications, however, it is often more appropriate to interpret $A$ as a \defn{source} and $B$ as a \defn{target}.

\paragraph{Committor probabilities.}

Associated to the forward and backward reactivities are the \defn{forward} and \defn{backward committors} $q_{i}^{\pm}(n)$ defined for $i \in S$ by
\begin{equation} \label{eq:committor-definition}
    q_{i}^{\pm}(n) := \mathbb P (R^{\pm}(n) \mid X_n = i) .
\end{equation}
Consider an ensemble of trajectories such that each trajectory satisfies $X_0 = i$. Only a certain fraction of these trajectories will reach set $B$ before set $A$; this fraction is precisely $q_i^+(n)$. States for which $q_i^+(n)$ takes large values are typically ``close’’ to $B$, in the sense that there exists at least one relatively short path from $i$ to $B$, although this need not hold in general. An analogous interpretation applies to $q_i^-(n)$ for the time-reversed Markov chain. The committors constitute the central objects of TPT, as they encode all information about the (infinite) past and future of the process relevant to transitions between $A$ and $B$.

A straightforward application of first-step analysis \citep{Helfmann-etal-20} shows that the committors are independent of $n$ and satisfy linear matrix equations
\begin{equation}
    q^{+}_i = 
    \begin{cases}
        \sum_{j \in S} P_{ij} q_{j}^{+} & \text{if }i \notin A \bigcup B,\\ 
        0 & \text{if }i \in A,\\ 
        1 & \text{if }i \in B,
    \end{cases}
    \quad \text{and} \quad   
    q^{-}_i = 
    \begin{cases}
        \sum_{j \in S} P_{ij}^{-} q_{j}^{-} & \text{if }i \notin A \bigcup B,\\ 
        1 & \text{if }i \in A,\\ 
        0 & \text{if }i \in B.
    \end{cases}
\end{equation}

\paragraph{TPT statistics.}

Using the committors, a number of statistics can be computed for reactive trajectories: the density, the current, the departure rate, the duration, and the remaining duration of reactive trajectories. The first four are discussed in \cite{Helfmann-etal-20}, while the fifth was introduced in \cite{Bonner-etal-23}.

\begin{itemize}
   \item The \defn{reactive density} is defined by
    \begin{equation} \label{eq:muAB}
        \mu^{AB}_{i}(n) := \mathbb P(X_n = i,\, R(n)) = q_{i}^- \pi_i q_{i}^+, \quad i\notin A\bigcup B.
    \end{equation}
    States with large reactive densities relative to their neighbors are interpreted as bottlenecks for reactive trajectories. 

    \item We also have the \defn{reactive current}, given by
    \begin{equation}
        f^{AB}_{ij}(n) := \mathbb P(X_n = i,\, X_{n+1} = j,\, R(n)) = q_{i}^{-} \pi_i P_{ij} q_{j}^+, \quad i, j \in S,
    \end{equation}
    as well as the \defn{effective reactive current}
   \begin{equation} \label{eq:effective-reactive-current}
       f_{i j}^{+} := \max\big\{f^{AB}_{ij} - f^{AB}_{j i}, 0\big\}.
   \end{equation}
   The effective reactive current is a $B$-facing gradient of the reactive density; it identifies pairs of states with a large net flow of probability. 

   \item The \defn{departure rate} from $i\in A$ is defined as the probability of a reactive trajectory to leave $i \in A$, and can be computed by summing up the reactive current that exits $i\in A$:
   \begin{equation}
      k^{i\in A\to}(n) := \mathbb P(X_n = i,\, R(n)) = \sum_{j\in S} f^{AB}_{ij},\,i\in A.
   \end{equation}
   Similarly, the \defn{arrival time} to $i\in B$ is represents the probability of a reactive trajectory to reach $i \in A$, and can be computed by summing up the reactive current that enters $i\in B$:
   \begin{equation}
      k^{i\in B\leftarrow}(n) := \mathbb P(X_n = i,\, R(n)) = \sum_{j\in S} f^{AB}_{ji},\,i\in B.
   \end{equation}

   \item The \defn{duration} of a reactive trajectory connecting $A$ with $B$, $t^{AB}$, in its original definition \citep{VandenEijnden-06} for diffusion processes in continuous time is given by the limiting ratio of the time spent during reactive transitions from $A$ to $B$ to the rate of reactive transitions leaving $A$. In  \cite{Helfmann-etal-20}, the following expression is provided in the case of discrete-time Markov chains:
   \begin{equation} \label{eq:tAB-vanE}
       t^{AB}(n) := \frac{\mathbb P(R(n))}{\mathbb P(X_n \in A, R(n))} = \frac{\sum_{i \in S} \mu_{i }^{AB}}{\sum_{i \in A, j \in S} f_{i j}^{AB}}.
   \end{equation}

   \item In \cite{Bonner-etal-23} a generalization of \eqref{eq:tAB-vanE} is introduced, which allows one to express it as a straightforward expectation. The aim of this generalization is to provide local information, that is, information about reactive trajectories at a particular state that have already left $A$. Let
   \begin{equation}
       E := \Big\{i \notin B :  \sum_{j \in S }P_{ij} q_j^+ > 0 \Big\}.
   \end{equation}
   The \defn{remaining duration} $t^{iB}$ is defined as
   \begin{equation} \label{eq:trem_def}
       t^{i\mathbb B}(n) 
       :=
       \begin{cases}
           \mathbb E\big[\tau_B^+(n) \mid X_n = i,R^+(n)\big] & \text{if }i \notin B,\\ 0 & \text{if }i \in B.
       \end{cases}
    \end{equation}
    A similar quantity is referred to as the lead time in \cite{Finkel-etal-21}. In \cite{Bonner-etal-23}, it is shown that remaining duration is independent of $n$ and satisfies the following system of linear equations:
    \begin{equation} 
        t^{iB} = 
       \begin{cases}
           1 + \sum_{j \in E}\frac{P_{ij} q_j^+}{\sum_{k \in S}P_{ik} q_k^+} t^{jB} & \text{if }i \in E,\\ 0 & \text{if }i \in B.
       \end{cases}
    \label{eq:trem_comp}
    \end{equation}
    When $A$ contains only one state, we also have that
    \begin{equation}
        t^{iB} \big|_{i = A} = t^{AB} + 1,
    \end{equation}
    where $t^{AB}$ is defined in \eqref{eq:tAB-vanE}.

\end{itemize}

\subsection{Open systems and connectivity}

When trajectory data are available only on an open domain, an appropriate preprocessing step is required in order to construct a well-defined Markov chain. Following \cite{Miron-etal-21-Chaos} and subsequent related studies, we introduce a \defn{two-way nirvana state}, denoted by $\omega$, to render the system closed. Assume that all trajectory data lie within a subdomain $s \subset S$. We decompose $s$ as $s = S^O \bigcup \omega$ such that $\partial s \subset \omega$ and $|\omega| \ll |S^O|$. Next, we construct a covering of $S^O$ by $N$ boxes and append one additional box representing the entirety of $\omega$. By counting transitions between these boxes, we obtain a row-stochastic transition matrix of the form
\begin{equation}
    P = 
    \begin{pmatrix}
        P^{O \to O} & \mathbf p^{O \to \omega} \\ 
        \mathbf p^{\omega \to O} & 0 
    \end{pmatrix},
    \label{eq:Pclose}
\end{equation}
where $\smash{P^{O \to O}} \in \smash{\mathbb R^{N \times N}}$, $\smash{\mathbf p^{O \to \omega}} \in \smash{\mathbb R^{N \times 1}}$, and $\smash{\mathbf p^{\omega \to O}} \in \smash{\mathbb R^{1\times N}}$. Note that trajectories that both originate and terminate in the nirvana state are excluded from consideration. More generally, we focus exclusively on reactive trajectories that do not visit this auxiliary nirvana state. Imposing this constraint is equivalent to replacing $A$ by $A \bigcup \omega$ and $B$ by $B \bigcup \omega$ in the computation of $\mathbf q^+$ and $\mathbf q^-$, respectively. It can be shown \cite{Miron-etal-21-Chaos} that this procedure is also equivalent to leaving $A$ and $B$ unchanged while instead replacing $P$ with the row-substochastic matrix $P^{O \to O}$ and $\boldsymbol\pi$ with the restriction of the stationary distribution of \eqref{eq:Pclose} to the set $O$. 

Depending on the empirical structure of the data, the transition matrix $P^{O \to O}$ may fail to be irreducible and aperiodic. To address this issue, one can apply Tarjan's algorithm \cite{Tarjan-72} to identify and extract the largest strongly connected component of $P^{O \to O}$. The matrix $P^{O \to O}$ is then modified by eliminating all states outside this component, as well as all transition contributions arising from trajectories that traverse the removed states. The resulting transition matrix is irreducible and aperiodic, excludes the nirvana state, and is therefore suitable for use in the TPT formulation described above.

\setcounter{equation}{0}
\section{Bayesian inversion}

Our objective here is to determine the subset of $A$ that serves as the origin of the trajectories that are observed to enter $B$ at time $t^B > 0$. Unlike in TPT, we do not impose constraints on the way trajectories connect $A$ and $B$. 

Motivated by the analysis presented in \cite{Miron-etal-19-Chaos}, we consider the computation of the probability that the Markov chain lies in $B$ at time $n \ge 0$, conditional on the initial state being $a \in A$. This is achieved by propagating forward a probability vector that is initially supported on the single state $a$, $\mathbf 1_a$, while at each step recording the probability in $B$:
\begin{equation}
    p_{B,a}(n) := \mathbb P(X_n \in B \mid X_0 = a) = \mathbf 1_a P^n\cdot \mathbf 1_B = \sum_{i\in B}(P^n)_{ia}.
    \label{eq:p}
\end{equation}
Provided that $B$ is \emph{absorbing}, it follows that
\begin{equation}
    p_{B,a}(n) = \mathbb P(\tau_B^+ \leq n\mid X_0 = a). 
\end{equation}
This is because, by definition, each trajectory that visits $B$ has a first visit at some time $n$ and subsequently remains in $B$. Since the first visit events $\{\tau^+_B = n\}_{n\ge 0}$ are mutually exclusive, it follows that
\begin{equation}
    p_{B,a}(n) = \sum_{k = 0}^n \mathbb P(\tau^+_B = k\mid X_0 = a). 
\end{equation}
Then, by replacing $n$ with $n - 1$ and subtracting each equation, we obtain:
\begin{equation}
    p(n\mid a) := \mathbb P(\tau^+_B = n\mid X_0 = a) = \begin{cases} p_{B,a}(n) = 0 & \text{if } n = 0,\\ p_{B,a}(n) - p_{B,a}(n-1) & \text{if } n > 0.\end{cases}
    \label{eq:hit-prob}
\end{equation} 

We consider $B$ and $t^B$ as \emph{observations} of the random variables $X_*$ and $\tau_*$. Since $B$ is absorbing, it follows that $\tau^+_B < \infty$ and $X_* = B$, by \eqref{eq:first-passage-forward}. Consequently, the events $\{X_* = B,\,\tau_* = t^B\}$ and $\{\tau_B^+ = t^B\}$ are equivalent. Therefore, conditional on the chain being in $a$ at time $n=0$, the probability that the random variables jointly assume their observed values
\begin{equation}
     \mathbb P(X_* = B,\,\tau_* = t^B\mid X_0 = a) \equiv \mathbb P(\tau^+_B = t^B\mid X_0 = a) = p(t^B\mid a),
\end{equation}
which can be computed using \eqref{eq:hit-prob}. 

Bayesian inversion \cite{Bolstad-Curran-16} provides a probabilistic depiction of the origin ($a$) of trajectories based on their observed location ($B$) and time ($t^B$). By Bayes' theorem, the \defn{posterior distribution} of $a$, or the probability distribution of $a$ after observing $t^B$, is computed as
\begin{equation}
  p(a\mid t^B) \propto p(t^B\mid a)\, p(a),
  \label{eq:post}
\end{equation}
where $p(a)$ denotes the \defn{prior distribution} of $a$, which represents its existing knowledge before observing anything. Finally, \defn{maximum likelihood estimator} of $a$ is given by
\begin{equation}
    \hat a := \underset{a}{\operatorname{arg\,min}}\,p(a\mid t^B).    
\end{equation}

\bibliographystyle{abbrvnat}
%\bibliography{fot}

\end{document}